\documentclass[12pt,preprint,preprintnumbers,a4paper]{article}
\pdfoutput=1

\usepackage{jheppub}

\usepackage{afterpage}

\usepackage{longtable}
\usepackage{multirow}
\usepackage{enumerate}
\usepackage{amsmath}
\usepackage{amsfonts}
\usepackage{amssymb}
\usepackage{graphicx, rotating}
\usepackage{epstopdf}
\usepackage{epsfig}
\usepackage{latexsym}
\usepackage{graphicx}
\usepackage{color}
\usepackage{amsmath,amssymb}
\usepackage{slashed}
\usepackage{hyperref}
\usepackage{wasysym}
\usepackage{MnSymbol}
\usepackage{caption}

\definecolor{LightCyan}{rgb}{0.88,1,1}
\definecolor{LightViolet}{rgb}{1, 0.88,1}
\definecolor{LightGreen}{rgb}{0.88, 1, 0.88}
\definecolor{forestgreen}{rgb}{0.13,0.35,0.13}

\usepackage{color, colortbl}
\definecolor{rossos}{cmyk}{0,1,1,0.55}
\definecolor{bluscuro}{rgb}{0.15, 0.2, .85}
\definecolor{bluchiaro}{cmyk}{1,.3,0.,0.1}
\definecolor{verdechiaro}{rgb}{0.6,.1,0.9}
\definecolor{forestgreen}{rgb}{0.13,0.35,0.13}
\definecolor{gold}{rgb}{1,0.84,0}
\definecolor{darkcyan}{rgb}{0,0.55,0.55}
\definecolor{brown}{rgb}{0.65,0.16,0.16}
\definecolor{orange}{rgb}{1,0.65,0.}

\def\bea{\begin{eqnarray}} \def\eea{\end{eqnarray}}
\def\be{\begin{equation}} \def\ee{\end{equation}}

\newcommand{\promille}{%
  \relax\ifmmode\promillezeichen
        \else\leavevmode\(\mathsurround=0pt\promillezeichen\)\fi}
\newcommand{\promillezeichen}{%
  \kern-.05em%
  \raise.5ex\hbox{\the\scriptfont0 0}%
  \kern-.15em/\kern-.15em%
  \lower.25ex\hbox{\the\scriptfont0 00}}

 \hypersetup{colorlinks, citecolor=bluscuro, linkcolor=bluscuro, urlcolor=verdechiaro}

\begin{document}
\preprint{MIT-CTP/4624}

\vspace*{-30mm}

\title{\boldmath Constraining the Higgs portal\\ with antiprotons}

\author[a]{Alfredo Urbano}
\author[b,c,a]{and Wei Xue}
\affiliation[a]{SISSA - International School for Advanced Studies, via Bonomea 265, I-34136, Trieste, ITALY.}
\affiliation[b]{Center for Theoretical Physics, Massachusetts Institute of Technology, Cambridge, MA 02139, USA.}
\affiliation[c]{INFN sez. di Trieste, via Bonomea 265, I-34136, Trieste, ITALY.}

\emailAdd{alfredo.urbano@sissa.it}
\emailAdd{weixue@mit.edu}

\vspace{2cm}


\abstract{
The scalar Higgs portal is a compelling model of dark matter (DM) in which a renormalizable coupling with the Higgs boson
provides the connection between the visible world and the dark sector.
In this paper we investigate the constraint placed on the parameter space of this model by the antiproton 
data. Due to the fact that the antiproton-to-proton ratio has relative less systematic uncertainties than 
the antiproton absolute flux, we propose and explore the possibility to combine all the available $\bar{p}/p$ data. 
Following this approach, we are able to obtain stronger limits if compared with the existing literature. 
In particular, we show that 
most of the parameter space close to the Higgs resonance is ruled out by our analysis.
Furthermore, by studying the reach of the future AMS-02 antiproton and 
antideuteron data, we 
argue that a DM mass of $\mathcal{O}(150)$ GeV offers a promising discovery potential. 
The method of combining all the antiproton-to-proton ratio data proposed in this paper is quite general, and can be straightforwardly applied to other models.
}
\maketitle

\section{Introduction}
Since the discovery of the first \textit{negative proton} in 1955 \cite{Segre}, 
antiprotons have become a fundamental pillar in experimental high-energy physics, both in collider physics and astrophysics.
In particular, antiprotons play a starring role in the context of indirect detection of dark matter (DM):
they are copiously produced 
in the final stages of DM annihilations into Standard Model (SM) particles
as a consequence of showering and hadronization processes. This is in particular true considering DM annihilation into hadronic channels, 
like for instance annihilation into $\bar{b}b$. However, thanks to the electroweak radiative corrections, 
this is also true for DM annihilations into leptonic channels, providing that the DM mass is around the TeV scale \cite{Kachelriess:2009zy,Ciafaloni:2010qr,Ciafaloni:2010ti}.

Furthermore, the astrophysical background plaguing this potential DM signal is relatively well understood. 
In a standard scenario it mostly consists in secondary antiprotons originated from 
 the interactions of primary cosmic-ray protons, produced in supernova remnants, with the interstellar gas.
 For these reasons the antiproton channel is considered one of the most promising 
probes to shed light on the true nature of DM \cite{Bergstrom:1999jc,Barrau:2005au,Chardonnet:1996ca,Bergstrom:2006tk}. 

However, all the experimental data collected so far show a fairly good agreement with the predictions of the astrophysical background, usually computed by means of dedicated codes such as \texttt{GALPROP} or \texttt{DRAGON}. 
Overturning the previous perspective, this negative results is often exploited to 
place strong bounds on the annihilation cross-section of DM into SM 
particles (see, e.g., refs.~\cite{Fornengo:2013xda,Evoli:2011id,Cirelli:2013hv,Asano:2012zv,Cheung:2010ua,Cheung:2010fu,Garny:2011ii,Cerdeno:2011tf,Lavalle:2011fm,DeSimone:2013fia,Cirelli:2014lwa,Bringmann:2014lpa}).
Following this line, in this paper we explore the constraining power of the antiproton data considering as a benchmark example the so-called Higgs portal DM model \cite{Silveira:1985rk,McDonald,Burgess,PW}. 
Despite its simplicity, in fact, this model offers a rich phenomenology, and it provides a simple and motivated  paradigm of DM.

The aim of this paper is twofold.
On the one hand, we extract our bounds focusing the analysis on the experimental data
describing the antiproton-to-proton ratio instead of (as customary in the literature) the absolute antiproton flux. In this way we can get rid of 
the systematic uncertainties  that usually preclude the comparison between data taken by different experiments.
On the other one, we compare, in the context of the Higgs portal model and in a wide range of DM mass, 
our results with the bounds obtained considering  the invisible Higgs decay width and the spin-independent 
DM-nucleon cross-section. The purpose of this comparison is to highlight the regions of the parameter space in which the antiproton data give the most stringent limits.

This paper is organized as follows. In section~\ref{sec:HiggsPortal}, we briefly introduce the scalar Higgs portal model.
In section~\ref{sec:Analysis}, we discuss all the relevant aspects of our analysis; in particular, we present in detail 
the computation of the $\bar{p}/p$ flux considering both the standard astrophysical background and the DM signal.
In section~\ref{sec:Results} we present our results, and in section~\ref{sec:future} we discuss future prospects.
Finally, we conclude in section~\ref{sec:Conclusions}. In appendix~\ref{sec:FermionHiggsPortal}, we generalize our results to the fermionic Higgs portal model.

\section{Setup: the scalar Higgs portal}\label{sec:HiggsPortal}

The Lagrangian of the scalar Higgs portal  model that we consider in this work
 has the following structure
\cite{Silveira:1985rk,McDonald,Burgess,PW}
\begin{equation}\label{eq:HiggsPortal}
\mathcal{L}_{\rm HP} = \mathcal{L}_{\rm SM}
+ \frac{1}{2}(\partial_{\mu}S)(\partial^{\mu}S)-\frac{m_{0}^2}{2}S^2
-\frac{\lambda_{\rm S}}{2}|H|^2S^2~,
\end{equation}
where $\mathcal{L}_{\rm SM}$ is the SM Lagrangian, $\lambda_{\rm S}$ is the Higgs portal coupling,
and the real field $S$ is a scalar gauge singlet with mass -- after electroweak symmetry breaking -- given by $m_{\rm S}=(m_{0}^2 + \lambda_{\rm S}v^2/2)^{1/2}$; $H$ is the SM Higgs doublet with vacuum expectation value (vev) $\langle H\rangle = v/\sqrt{2}=174$ GeV.
 
 The relevant parameter space of the model is the two-dimensional plane $(m_{\rm S},\lambda_{\rm S})$.
 After electroweak symmetry breaking the Higgs portal coupling generates the trilinear vertex $\mathcal{L}_{\rm hS^2}=
 \lambda_{\rm S}vhS^2/2$, where $h$ is the physical Higgs boson; this interaction is responsible for all the phenomenological properties of the model since via this vertex the DM particle communicates with all the SM species. In this paper, we focus on the possibility that the scalar field $S$ plays the role of cold DM in the Universe. 
 
 In appendix~\ref{sec:FermionHiggsPortal}, we will analyze a different type of Higgs portal with fermionic DM.

\subsection{Relic density}\label{sec:RelicDensity}

Through the exchange of the Higgs in the s-channel, two DM particles can annihilate into all the SM final states that are
kinematically allowed by the value of the DM mass, $m_{\rm S}$. The annihilation cross-section 
times the relative velocity  $v_{\rm rel}$ of the two annihilating DM particles
takes the remarkably simple form 
 \begin{equation}\label{eq:Xsection}
\sigma v_{\rm rel}=\frac{2}{\sqrt{s}}\left[\frac{\lambda_{\rm S}^2 v^2}{(s-m_h^2)^2+\Gamma_{\rm h,S}^2 m_h^2}\right]\Gamma_{\rm h}(\sqrt{s})~,
\end{equation}
where the square of the total energy in the c.o.m. frame is $s=4m_{\rm S}^2/(1-v_{\rm rel}^2/4)$. In eq.~(\ref{eq:Xsection})
$\Gamma_{\rm h}(\sqrt{s})$ is the off-shell decay width of the Higgs boson (with $m_h^*=\sqrt{s}$), summed over all the  SM final states. We use the public code HDACAY \cite{Djouadi:1997yw}
  to compute the width $\Gamma_{\rm h}(\sqrt{s})$. In this way we are able to include \textit{i})
  the $\mathcal{O}(\alpha_s)$ NLO QCD radiative corrections to the Higgs decay into quarks  
  and \textit{ii}) the Higgs decay modes into off-shell gauge bosons. The importance of these radiative effects has been emphasized in ref.~\cite{Cline:2012hg}.
\begin{figure*}[!htb!]
\minipage{0.5\textwidth}
  \includegraphics[width=\linewidth]{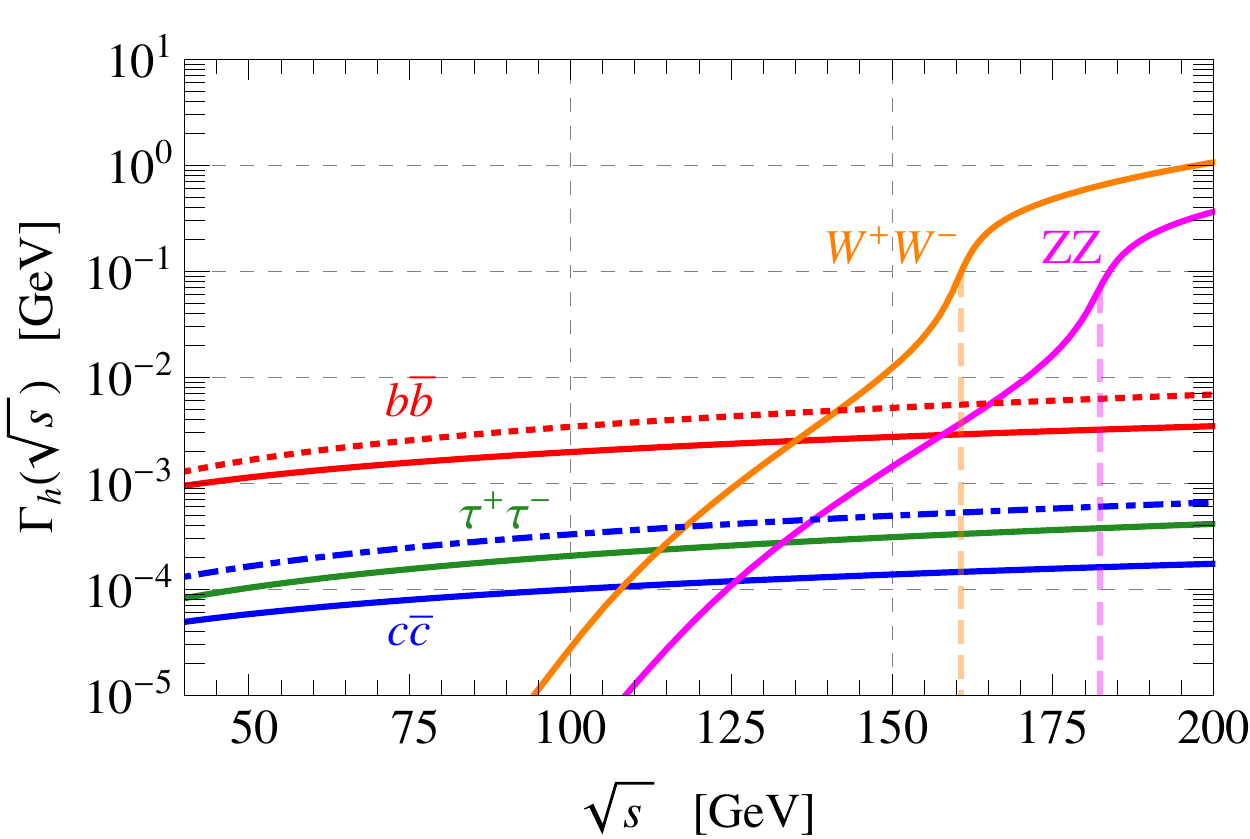}
\endminipage
\minipage{0.5\textwidth}
  \includegraphics[width=\linewidth]{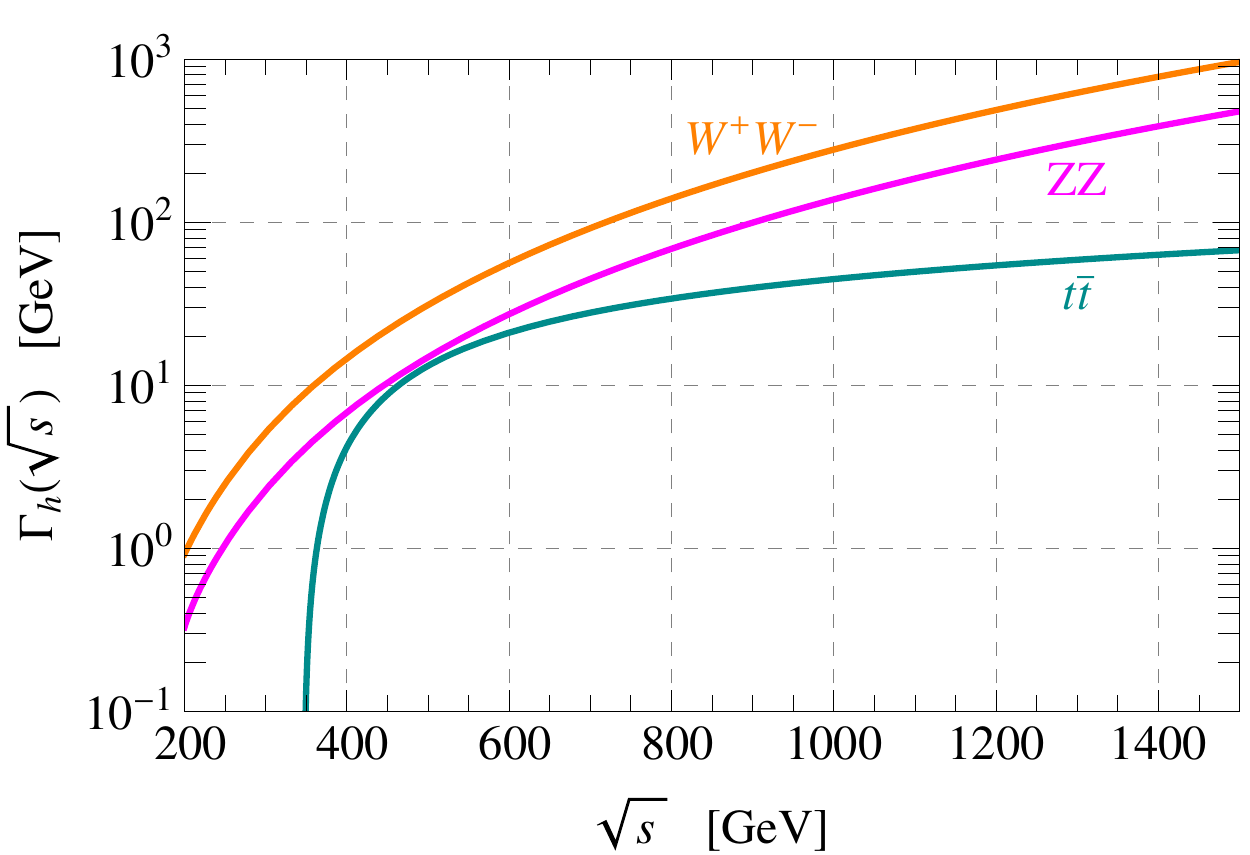}
\endminipage
 \caption{\textit{
 Off-shell decay width of the Higgs boson $\Gamma_{\rm h}(\sqrt{s})$ in the energy interval $\sqrt{s}\in[40,200]$ GeV obtained using the public code HDECAY (left panel, solid lines), including the NLO QCD corrections to the Higgs decay into quarks  and the Higgs decay modes into off-shell gauge bosons. We show separately the most important contributions relevant for the computation of the annihilation cross-section in eq.~(\ref{eq:Xsection}). The impact of the radiative corrections 
 is proven by the comparison with the corresponding tree-level expressions (tree-level Higgs decay width into $b\bar{b}$, dotted line, and $c\bar{c}$, dot-dashed line); we also show the $W^+W^-$ and $ZZ$ kinematical thresholds (vertical dashed lines) to emphasize the importance of the Higgs decay modes into off-shell gauge bosons in the region close to the Higgs resonance, $\sqrt{s}=126$ GeV. At higher energies, $\sqrt{s}> 200$ GeV, we compute
 $\Gamma_{\rm h}(\sqrt{s})$ analytically (right panel).   
 }}\label{fig:HiggsGamma}
\end{figure*}
We plot in the left panel of fig.~\ref{fig:HiggsGamma} the function $\Gamma_{\rm h}(\sqrt{s})$ in the energy interval $\sqrt{s}\in[40,200]$ GeV (left panel); we separate 
the most important contributions, namely the Higgs decays into SM quarks ($b\bar{b}$ and $c\bar{c}$) and tau leptons as well as electroweak gauge bosons ($W^+W^-$ and $ZZ$). 
The importance of the radiative corrections is evident from the comparison with the corresponding tree level expressions (dashed lines in  fig.~\ref{fig:HiggsGamma}).

Going towards higher energies, there is an important issue to keep in mind.  In the SM the Higgs quartic coupling $\lambda$ is a function of the Higgs mass, i.e. $\lambda(m_h)=m_h^2/2v^2$ and, as a consequence, $\lambda(126)\simeq 0.13$. In the computation of the off-shell decay width $\Gamma_{\rm h}(\sqrt{s})$, some electroweak corrections involving the quartic coupling are overestimated at high energy, growing like $\lambda(\sqrt{s})=s/2v^2$. In order to get rid of this issue, for $\sqrt{s}>200$ GeV we compute $\Gamma_{\rm h}(\sqrt{s})$ analytically; we show the corresponding values for annihilation into $W^+W^-$, $ZZ$ and $t\bar{t}$ in the right panel of fig.~\ref{fig:HiggsGamma} in the energy interval $\sqrt{s}\in[200, 1500]$ GeV. Finally, notice that above the kinematical threshold $\sqrt{s}>2m_h=252$ GeV DM annihilation into two Higgses is kinematically allowed; the corresponding annihilation cross-section, however, cannot be recast in the form described by eq.~(\ref{eq:Xsection}) since, in addition to the s-channel exchange of the Higgs, also t- and u-channel diagrams in which the DM particle is exchanged contribute to the amplitude. We include this channel computing the cross-section analytically (see, e.g., ref.~\cite{Guo:2010hq}).

 In eq.~(\ref{eq:Xsection}) $\Gamma_{\rm h,S}$
represents the on-shell decay width of the Higgs boson and  it consists of two pieces, namely $\Gamma_{\rm h,S}\equiv \Gamma_{\rm h}^{\rm SM}+\Gamma_{\rm h\to SS}$; $ \Gamma_{\rm h}^{\rm SM}=4.217$ MeV is the SM contribution while $\Gamma_{\rm h\to SS}$ is the decay width describing the process $h\to SS$, kinematically allowed if $m_{\rm S}<m_h/2$.
The explicit expression of $\Gamma_{\rm h\to SS}$ is discussed in the context of the LHC
 bound (see section~\ref{sec:LHCbound}, eq.~(\ref{eq:InvisibleDecayWidth})).

The thermally averaged annihilation cross-section is given by
  \begin{equation}
  \langle \sigma v_{\rm rel} \rangle =
  \int_{4m_{\rm S}^2}^{\infty}ds\, \frac{s\sqrt{s-4m_{\rm S}^2}
  K_1(\sqrt{s}/T)}{16Tm_{\rm S}^4K_2^2(m_{\rm S}/T)}\,\sigma v_{\rm rel}~,
  \end{equation}
 where $T$ is the temperature and $K_{\alpha=1,2}$ are the modified Bessel functions of second kind. We numerically solve the Boltzmann equation describing the evolution of the number density $n(x)$ of the DM particles during the expansion of the Universe, being $x\equiv m_{\rm S}/T$. In terms of the yield ${\rm Y}(x)=n(x)/s(x)$, where $s(x)$ is the entropy density, this equation reads 
  \begin{equation}
  \frac{d{\rm Y}}{dx}=-Z(x)\left[{\rm Y}^2(x) - {\rm Y}_{\rm eq}^2(x)\right]~,
  \end{equation}
  where
  \begin{equation}
  Z(x) \equiv \sqrt{\frac{\pi}{45}}\frac{m_{\rm S}M_{\rm PL}}{x^2}\sqrt{g_*(T)}
   \langle \sigma v_{\rm rel} \rangle(x)~.
  \end{equation}
 $M_{\rm PL}=1.22\times10^{19}$ GeV is the Planck mass, and $g_*(T)$ is the number of relativistic degrees of freedom. At the equilibrium, we have
  \begin{equation}\label{eq:EquilibriumYield}
  {\rm Y}_{\rm eq}(x) = \frac{45}{4\pi^4}\frac{x^2}{h_{\rm eff}(T)}K_2(x)~,
  \end{equation}
  where $h_{\rm eff}(T)$ is the effective entropy.\footnote{We include in our numerical analysis the temperature dependence in $g_*(T)$ and $h_{\rm eff}(T)$; we extract the corresponding functions from the DarkSUSY code \cite{Gondolo:2004sc, DarkSUSYweb}.} The integration of the Boltzmann equation gives the yield today, ${\rm Y}_0$, which is related to the DM relic density through
  \begin{equation}\label{eq:RDtheory}
  \Omega_{\rm DM} h^2 = \frac{2.74\times 10^8 m_{\rm S} {\rm Y}_0}{{\rm GeV}}~.
  \end{equation}
  The Planck collaboration has recently reported the value \cite{Ade:2013zuv} $\Omega_{\rm DM}h^2= 0.1199 \pm 0.0027$ ($68\%$ C.L.); we compute the relic density according to eq.~(\ref{eq:RDtheory}), and we impose to match the measured value.
 
Before proceeding, let us stress that the value of the relative velocity sets the position of the Higgs resonance during 
DM annihilation. From eq.~(\ref{eq:Xsection}) it follows that the pole of the Higgs propagator is given by $s=m_h^2$; at zero relative velocity, therefore, the resonant annihilation occurs at $m_{\rm S}=m_h/2=63$ GeV, while in general it occurs at $m_{\rm S}^2=m_h^2(1-v_{\rm rel}^2/4)/4$. Considering the annihilation in the early Universe -- i.e. taking for definiteness $v_{\rm rel}=1/2$ -- the resonance occurs at $m_{\rm S}\simeq 61$ GeV.

\subsection{LHC bound}\label{sec:LHCbound}

If $m_{\rm S}<m_h/2$, the Higgs boson can decay into two DM particles resulting 
in the possibility to have a sizable invisible decay channel. In this case the invisible decay width of the Higgs is
  \begin{equation}\label{eq:InvisibleDecayWidth}
\Gamma_{\rm h\to SS}(m_{\rm S},\lambda_{\rm S}) = \frac{v^2\lambda_{\rm S}^2}{32\pi m_h}
\sqrt{1-\frac{4m_{\rm S}^2}{m_h^2}}~.
\end{equation}
The invisible branching ratio
\begin{equation}\label{eq:InvisibleBR}
{\rm BR_{inv}}(m_{\rm S},\lambda_{\rm S})\equiv \frac{\Gamma_{\rm h\to SS}(m_{\rm S},\lambda_{\rm S})}{\Gamma_{\rm h}^{\rm SM}+\Gamma_{\rm h\to SS}(m_{\rm S},\lambda_{\rm S})}
\end{equation}
is severely constrained by the current searches at the LHC \cite{ATLASinv,CMSinv}, and ${\rm BR_{inv}}>22\%$ is excluded at 95\% C.L. \cite{Falkowski:2013dza}. Using eq.~(\ref{eq:InvisibleBR}) it is straightforward to translate this bound in the parameter space $(m_{\rm S}, \lambda_{\rm S})$,  and in section~\ref{sec:Results} we will include this constraint in our analysis. 
To give a quantitative idea of the size of 
the invisible branching ratio in eq.~(\ref{eq:InvisibleBR}), notice that for the benchmark values 
$m_{\rm S}=20$ GeV, $\lambda_{\rm S}=0.05$ we have ${\rm BR_{inv}}\simeq 73\%$. As soon as the invisible decay channel is kinematically allowed, therefore, it can easily dominate over the SM final states even for relatively small values of the Higgs portal coupling. Notice that invisible width in eq.~(\ref{eq:InvisibleDecayWidth}) is equal to zero for the resonant value $m_{\rm S}=m_h/2$. Therefore, it will be impossible to test this particular region using the bound on the invisible branching ratio.

 \subsection{Direct detection constraint}\label{sec:DD}
 
The Higgs portal interaction, through the exchange of the Higgs boson in the t-channel, provides the possibility to have a non-zero spin-independent elastic cross-section of DM on nuclei. 

Integrating out the Higgs in the limit of negligible exchanged momentum, it is possible to write the following effective interactions between DM and light quarks and gluons inside the nucleus
\begin{equation}\label{eq:DDS}
\mathcal{L}_{\rm S}^{\rm eff}= \frac{\lambda_{\rm S}}{2m_h^2}S^2\left(\sum_q m_q\overline{q}q
-\frac{\alpha_s}{4\pi}G^2\right)~,
\end{equation}
with $q=u,d,s$, and $G^2=G_{\mu\nu}G^{\mu\nu}$, where $G_{\mu\nu}$ is the gluon field strength. Using this effective interaction, it is straightforward to compute the spin-independent cross-section describing the elastic DM-nucleon scattering
\begin{equation}\label{eq:SIHIGGS}
\sigma_{\rm SI} = \frac{\lambda_{\rm S}^2 f_{\rm N}^2}{4\pi}\frac{\mu_{\rm S}^2 m_{\rm N}^2}{m_h^4 m_{\rm S}^2}~,
\end{equation}
where $\mu_{\rm S}\equiv m_{\rm N}m_{\rm S}/(m_{\rm N} + m_{\rm S})$ is the DM-nucleon reduced mass, $m_{\rm N} =0.946$ GeV is the
nucleon mass, and $f_{\rm N} = 0.303$ is the Higgs-nucleon coupling \cite{Cline:2012hg}. The LUX experiment has reported the strongest limit on $\sigma_{\rm SI}$  \cite{Akerib:2013tjd}. Using eq.~(\ref{eq:SIHIGGS}) we translate this bound in the parameter space $(m_{\rm S}, \lambda_{\rm S})$,  and in section~\ref{sec:Results} we will include this constraint in our analysis. 

Before proceeding, let us stress a simple but important point. The square of the momentum transferred
in a typical DM-nucleus elastic scattering always satisfies the condition $-q^2 \ll m_h^2$, with 
$q^2 = -2m_{\rm Xe}E_{\rm rec}$ where the mass of a Xenon nucleus is $m_{\rm Xe}= 121$ GeV and for the typical recoil energy one has $E_{\rm rec} \sim $ few keV. 
This simply implies that there is no resonant enhancement in elastic scatterings via the Higgs portal; as a consequence, the region with $m_{\rm S}\approx 63$ GeV where the model reproduces the correct relic abundance is beyond the present reach of direct detection experiments, 
and it will be covered only in the next future.

\section{Antiprotons: Background vs Signal}\label{sec:Analysis}

In this section we address the properties of the transport equation 
describing the propagation of cosmic rays in the Galaxy. In section~\ref{sec:Method}, 
we discuss the background contribution, focusing our attention mostly on 
astrophysical background of protons and antiprotons. 
In section~\ref{sec:BGvsSignal}, we illustrate the strategy that we follow in order to 
extract our bound on the scalar Higgs portal model.

\subsection{Selection of propagation models}\label{sec:Method}

Considering the total luminosity injected  in the Milky Way galaxy via cosmic rays, 
 most ($\sim 90$\%) of it consists of primary protons, $\sim 10$\% of helium nuclei, 
a further $\sim 1$\% of heavier nuclei, and $\sim 1$\% of free electrons.
In full generality, the evolution of the cosmic-ray density in the Galaxy is described by the following 
transport equation 
\begin{eqnarray}
   \frac{ \partial N_i} { \partial t} &=&
      \vec{\nabla} \cdot \left( D \vec{\nabla} - \vec{v_c} \right) N_i
      + \frac{\partial} { \partial p} \left ( \dot{p} 
      -\frac{p}{3} \vec{\nabla} \cdot \vec{v_c} \right) N_i  
   + \frac{ \partial} { \partial p} p^2 D_{pp} \frac{\partial}{\partial p}
      \frac{ N_i} { p^2}  +   Q_i ( p, r, z) 
   \nonumber \\
     &+&  \sum_{j>i} \beta n_{gas} (r,z) \sigma_{ji} N_j
      - \beta n_{gas} \sigma_i^{in}( E_k) N_i ~,
   \label{eqn::prop}
\end{eqnarray}
where $N_i = N_i ( r, z, p, t) $ is the number density per total unit momentum of the  
$i$-th atomic species, $p$ is its momentum and $\beta$ is its velocity. 
The construction of a propagation model consists in solving eq.~(\ref{eqn::prop}) with a certain
boundary condition for all the cosmic-ray species; in this way one can compute -- 
for a given distribution and energy spectrum of the sources -- 
the spatial distribution and energy spectrum after propagation. 
Eq.~(\ref{eqn::prop}) contains a number of free parameters to be determined. 
Let us discuss these parameters one by one.
   \begin{itemize}

\item For each nucleus, the source distribution and the injection index $\gamma_i$ (or indices, if a break is considered).  In eq.~(\ref{eqn::prop})
these informations are encoded into the source term $Q_i ( p, r, z)$. Considering the contribution of the astrophysical background, 
$Q_i ( p, r, z)$ describes the distribution and injection spectrum of supernova remnants \cite{Ferriere:2001rg}. As far as the spectral index is concerned, we assume -- following ref.~\cite{Evoli:2011id} -- the same spectral index $\gamma_i = \gamma$ for all the nuclear species.

\item The normalization and energy dependence of the diffusion coefficient, $D$. In eq.~(\ref{eqn::prop}), we assume the following functional form
\begin{equation}
   D ( \rho, r, z) = D_0  \beta^\eta  \mathrm{e}^{|z|/z_t} \left( \frac{ \rho} {\rho_0}
      \right)^\delta~,
\end{equation}
where $\rho  = p \beta  / ( Ze )$ is the rigidity of the nucleus of charge $Z$, $D_0$ is the absolute normalization at reference 
   rigidity $\rho_0 = 3~\mathrm{GV}$, $\delta$ is the diffusion spectral index related to the turbulence of the interstellar medium,
   and $z_t$ is the scale height that controls the vertical spatial dependence, which is assumed to be exponential; 
  the halo thickness 
   $z_h \equiv 2z_t$ is the height of the propagation halo where stochastic diffusion and re-acceleration take place.
    An additional parameter, $\eta$, controls the low-energy behavior of the diffusion coefficient. 
   
\item The Alfv\'en velocity, $v_{A}$. It parametrizes the efficiency of the stochastic re-acceleration mechanism. In eq.~(\ref{eqn::prop}) it enters in the explicit expression of the diffusion coefficient in momentum space, $D_{pp}$ \cite{Evoli:2011id}.
   
\item The convective velocity, $\vec{v_c}$. It is the velocity of the convective wind, if present, that may contribute to the escape of cosmic rays from the Galactic plane. The convective velocity is zero in the Galactic plane and linearly increasing with the vertical distance $z$ from the Galactic plane.

\item  For each nucleus, the scattering cross-sections on the interstellar medium gas. The distribution of gas in the interstellar medium concentrates 
in the disk of the Galaxy.  Its number density is denoted as $n_{gas}$, which is mainly constituted by
 atomic and molecular hydrogen and helium. There are two main effects. On the one hand, the $i$-th nucleus is generated by the nuclear species $j$ with  cross-section $\sigma_{ji}$; on the other one, the $i$-th nucleus is destroyed by scattering on interstellar medium gas with total inelastic cross-section $\sigma_i^{in}$.
\end{itemize}
   
The propagation of cosmic rays can be simplified, 
if one takes into account only the high-energy region ( $\gtrsim 10$ GeV). 
In this regime, diffusion and energy losses play an
important role while other effects, 
such as convection and re-acceleration, are negligible.
However, following this approach one is forced to neglect all the cosmic ray data at low-energy; as a consequence, the ability of constraining DM
models -- especially with relative low mass -- decreases dramatically. In order
to take advantage of all the data set, the propagation equation needs to be solved numerically without these approximation.
To achieve this result, we use the public code \texttt{DRAGON} \cite{Evoli:2008dv,Gaggero:2013rya}, and
our procedure goes as follows.

Following ref~\cite{Evoli:2011id}, we start from five different benchmark propagation models: KRA, KOL, CON, THK and THN. These propagation models are characterized by different halo height $z_t$,  slope of the diffusion coefficient $\delta$, spectral index $\gamma$, and gradient of the convection velocity $dv_c/dz$. We collect the corresponding values in table~\ref{tab:models}. KRA, KOL and CON are characterized by the same halo height ($z_t = 4$ kpc) but they describe differently turbulence effects and convection velocity. THK and THN, on the contrary, explore two extreme values for the halo height, namely $z_t = 10$ kpc and $z_t = 0.5$ kpc. Using these five benchmark propagation models, the intent is to capture a wide range of astrophysical uncertainties. For a more detailed discussion we refer the interest reader to ref.~\cite{Evoli:2011id}.

\begin{figure*}[t]
\minipage{0.5\textwidth}
  \includegraphics[width=1.1\linewidth]{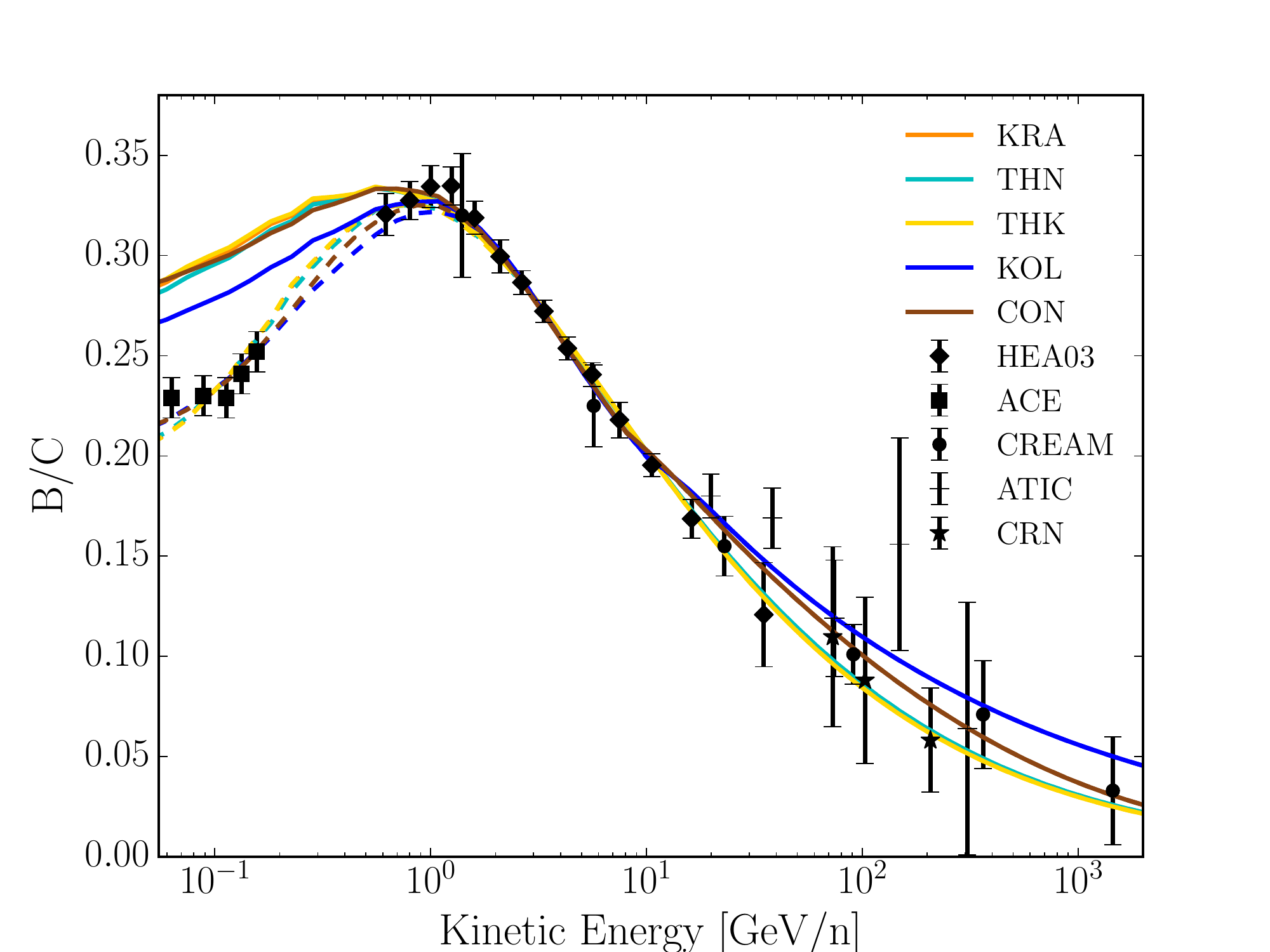}
\endminipage
\minipage{0.5\textwidth}
  \includegraphics[width=1.1\linewidth]{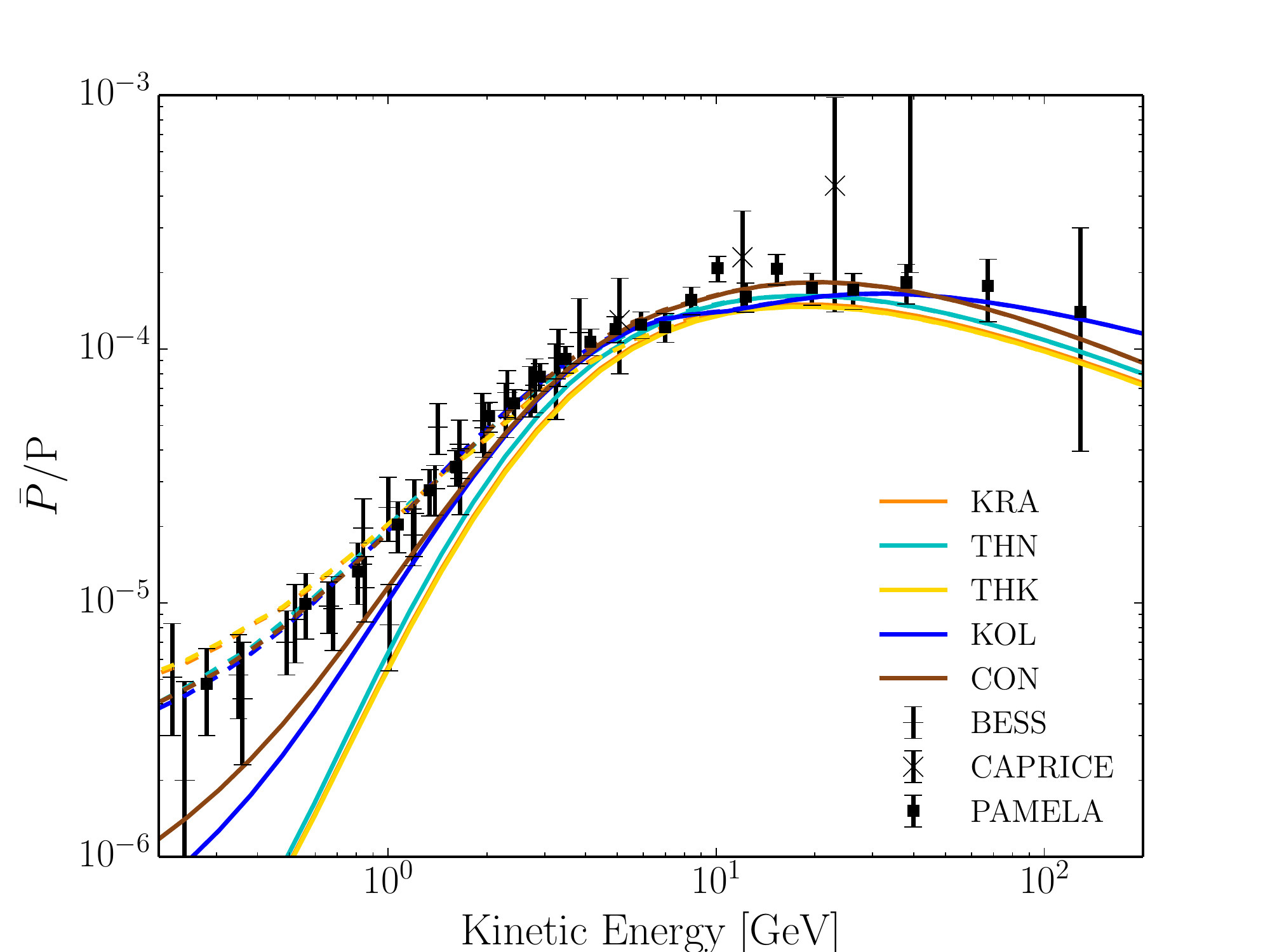}
\endminipage
 \caption{
\textit{
Fit of the B/C (left panel) and $\bar{p}/p$ (right panel) data 
using the five propagation models defined in table~\ref{tab:models}.
Solid lines correspond to the background model prediction without considering solar modulations, while dotted lines correspond to the prediction with solar modulations (see text for details).
}}
 \label{fig:5Background}
\end{figure*}

The second step in our analysis is to use the Boron-over-Carbon (B/C hereafter) data in order to determine -- for each one of the propagation setups defined before -- the remaining phenomenological parameters in 
eq.~(\ref{eqn::prop}). The B/C data employed in our analysis come from the HEAO3~\cite{Engelmann:1990zz}, CREAM~\cite{Ahn:2008my} 
ATIC~\cite{Panov:2007fe} and 
CRN~\cite{CRN} experiments. In table~\ref{tab:models}, we complete the definition of the five benchmark propagation models 
by minimizing the $\chi^2$ against B/C data; in this way we obtain, as output of the fitting procedure, $D_0$, $\eta$ and the Alfv\'en velocity $v_A$.
The B/C data with energy larger than 
$0.5$ GeV are considered, and solar modulation is fixed 
to the value $\Phi = 0.55$ GV. 
The reason why we chose to define our benchmark propagation models focusing exclusively on the B/C data 
is that antiproton
flux may originate both from astrophysical and exotic sources, while Boron and Carbon
 are usually generated only by astrophysical processes. Moreover, B/C represents the ratio between stable
secondary cosmic-ray flux divided by the corresponding primary cosmic-ray flux, which is exactly the 
same as $\bar{p}/p$ that we will use in the next step of our analysis. In the left panel of fig.~\ref{fig:5Background} we show the best-fit value for the B/C flux
for the five propagation models in table~\ref{tab:models}. 
Solid lines are obtained using $\Phi = 0$ GV, and the dashed line is 
to tune the value of $\Phi $ in order to fit  that low-energy data from the ACE collaboration. 
Notice that to compute $\chi^2$, we use fixed value of the solar modulation $\Phi= 0.55$ GV to 
fit the other experiments data.
 
Finally, we can use the five propagation models in table~\ref{tab:models} 
to compute the background contribution to the $\bar{p}/p$ flux. The choice of the  $\bar{p}/p$ ratio is made with the purpose of decreasing 
 the systematic uncertainties  that come from the comparison of data taken by different experiments.
 As far as the antiproton flux is concerned, for instance, various 
experiments can have different absolute flux due to different energy calibrations,
which we want to avoid. Using data describing the $\bar{p}/p$ flux, on the contrary, we can safely combine different datasets. 
   The $\bar{p}/p$ data employed in our analysis come from the BESS \cite{Orito:1999re,Asaoka:2001fv}, CAPRICE \cite{Boezio:2001ac} and PAMELA \cite{Adriani:2012paa} experiments. At this stage, the only free parameter is the value of solar modulation; in principle, since different experiments operated at different time, we can use  three independent values of $\Phi$ -- one for each experiment -- in order to fit the data, and in table~\ref{tab:models} we show the result of the corresponding $\chi^2$ fit (see caption for details).
\begin{table}[!htb!]
 \resizebox{0.68\textwidth}{!}{\begin{minipage}{\textwidth}
 \centering
\begin{tabular}[t]{|c||c|c|c|c||c|c|c|c||c|c|c|}
\hline
   \multirow{2}{*}{{\textbf{Model}}} & $z_t$&  \multirow{2}{*}{{$\delta$}} 
      &  \multirow{2}{*}{{$\gamma$}} & $dv_c/dz$ 
      & $D_0$ 
      & \multirow{2}{*}{{$\eta$}} & $v_A$ 
      & $\Phi$
      & \multirow{2}{*}{{$\chi^2_{\rm B/C}$}}
      & \multirow{2}{*}{{$\chi^2_{\bar{p}/p}$}}
      &  \multirow{2}{*}{{$\chi^2_{\bar{p}/p,\,{\rm PAMELA}}$}}   \\
      & (kpc)  &  & & (km s$^{-1}$ kpc$^{-1}$) 
      &  ($10^{28}$ cm$^2$ s$^{-1}$)
      &  
      &  (km s$^{-1}$) 
      & (GV)
      & 
      &  & \\  [1 pt] \hline\hline
   {\color{orange}{\textbf{KRA}}} & 4 &0.50 & 2.35 & 0 
      & 2.68 & -0.384 & 21.07  & 0.950
      & 0.95 & 1.26  & 1.08 \\
   \hline
   {\color{darkcyan}{\textbf{THN}}} & 0.5 &0.50 &  2.35 & 0 
      &  0.32 & -0.600 & 17.87 & 0.950
      & 0.88 & 1.41  & 1.26 \\
   \hline
   {\color{gold}{\textbf{THK}}} &  10 & 0.50 &  2.35 & 0
       & 4.45 & -0.332 & 19.91 & 0.950
      & 0.98 & 1.24  & 1.08 \\
   \hline
   {\color{blue}{\textbf{KOL}}} & 4 &0.33 & 1.78 / 2.45 & 0 
       & 4.45 & 1.00 & 40.00 & 0.673
      & 0.57 & 1.11  & 0.93 \\
   \hline
   {\color{brown}{\textbf{CON}}} &  4 & 0.60 & 1.93 / 2.35 & 50 
      &  0.99 & 0.786 & 40.00 & 0.19
      & 0.58 &  1.00 & 0.67 \\
   \hline
   \end{tabular}
   \end{minipage}}
   \caption{
   \textit{Phenomenological parameters describing the five benchmark propagation models (first column) used in our analysis.
   In the next four columns of the table we collect the values of halo height $z_t$, slope of the diffusion coefficient $\delta$, 
   spectral index $\gamma$, and gradient of the convective velocity $dv_c/dz$; these values are kept fixed, and define -- for each propagation models -- the
    corresponding properties of the diffusion-loss equation~(\ref{eqn::prop}). 
    The normalization of the diffusion coefficient $D_0$, the low-energy parameter $\eta$ and the Alfv\'en velocity $v_A$ are obtained 
    via a $\chi^2$ fit of the B/C data from the HEAO3~\cite{Engelmann:1990zz}, CREAM~\cite{Ahn:2008my},
ATIC~\cite{Panov:2007fe} and CRN~\cite{CRN} experiments; in addition, we show the corresponding minimum  $\chi^2_{\rm B/C}$ (divided by the number of degrees of freedom).
Using the propagation models so defined, we compute the background contribution to the $\bar{p}/p$ flux, and we fit the solar modulation potential $\Phi$ against data from the BESS, CAPRICE and PAMELA experiments. We show in the last two columns the corresponding values of $\chi^2_{\bar{p}/p}$
 (against, respectively, the full dataset and the subset of PAMELA data). The reported values for the solar modulation potential $\Phi$
 refer to the fit of the PAMELA data only.
   }
   }
\label{tab:models}
\end{table}

Once the astrophysical background has been fixed, we are now in the position to discuss the contribution from DM annihilation in the Higgs portal model.

\subsection{Antiproton bound on the scalar Higgs portal model}\label{sec:BGvsSignal}

We now move to discuss 
the computation of the antiproton flux originated from DM annihilation in the context of the Higgs portal model.

We solve eq.~(\ref{eqn::prop}) using as a source term
\begin{equation}\label{eq:DMsource}
Q_{\bar{p}}^{\rm DM}(p,r,z) = 
\frac{1}{2}
\left[
\frac{\rho_{\rm DM}(r,z)}{m_{\rm S}}
\right]^2
\left.\frac{dN}{dE}\right|_{\bar{p}}
\langle 
\sigma v_{\rm rel}
\rangle_0~,
\end{equation}
where $\rho_{\rm DM}$ is the DM density profile, $\left.dN/dE\right|_{\bar{p}}$ the antiproton emission spectrum, i.e. the number
of antiprotons per each annihilation, and $\langle 
\sigma v_{\rm rel}
\rangle_0$ is the thermally averaged annihilation cross-section times relative velocity evaluated at the present epoch. 
The solution of eq.~\ref{eqn::prop} allows us to compute -- as a function of DM mass and portal coupling, and at the location of the Earth -- the antiproton flux originated from DM annihilation, $\phi_{\bar{p}}^{\rm DM}(m_{\rm S},\lambda_{\rm S})$.
Next, we compute  the local $\bar{p}/p$ flux by combining DM contribution and astrophysical background, 
$\phi_{\bar{p}/p}(m_{\rm S},\lambda_{\rm S}) = [\phi_{\bar{p}}^{\rm BG} + \phi_{\bar{p}}^{\rm DM}(m_{\rm S},\lambda_{\rm S})]/\phi_{p}^{\rm BG}$. 
Finally, by means of a $\chi^2$ fit of the $\bar{p}/p$ data, we extract a bound on the parameter space of the scalar Higgs portal model.

We will discuss in section~\ref{sec:diffProf} the impact of different DM density profiles on the results of our analysis.
In the computation of the antiproton emission spectrum, we included -- consistently with the computation of  the relic density -- the three-body final states  consisting of one on-shell and one off-shell electroweak gauge bosons. 
Following ref.~\cite{Ciafaloni:2011sa}, we made use of the PYTHIA 8.1 event generator \cite{Sjostrand:2007gs,pythiaWS} to extract these energy spectra.

\section{Results}\label{sec:Results}

Following the approach outlined in section~\ref{sec:Analysis},
we derived the bound on the parameter space of the scalar Higgs portal model
by analyzing the antiproton-to-proton ratio data, and in this section we present and discuss our main results. 
We show the bound as a 3-$\sigma$ exclusion line in the planes $(m_{\rm S}, \lambda_{\rm S})$ and 
$(m_{\rm S}, \langle \sigma v_{\rm rel}\rangle_0)$. In both cases we compare the antiproton bound with the region that reproduces the correct amount of relic abundance, according to the result of the numerical analysis outlined in section~\ref{sec:RelicDensity}. In addition, we superimpose the constraints obtained considering the
invisible Higgs decay width
and the spin-independent DM-nucleon elastic cross-section as described, respectively, in sections~\ref{sec:LHCbound} and \ref{sec:DD}.

In section~\ref{sec:diffProp} we analyze the impact of different propagation models while in section~\ref{sec:diffProf} we discuss the impact of different DM density profiles. 
\subsection{On the impact of different propagation models}\label{sec:diffProp}

\begin{figure*}[!htb]
\minipage{0.5\textwidth}
  \includegraphics[width=\linewidth]{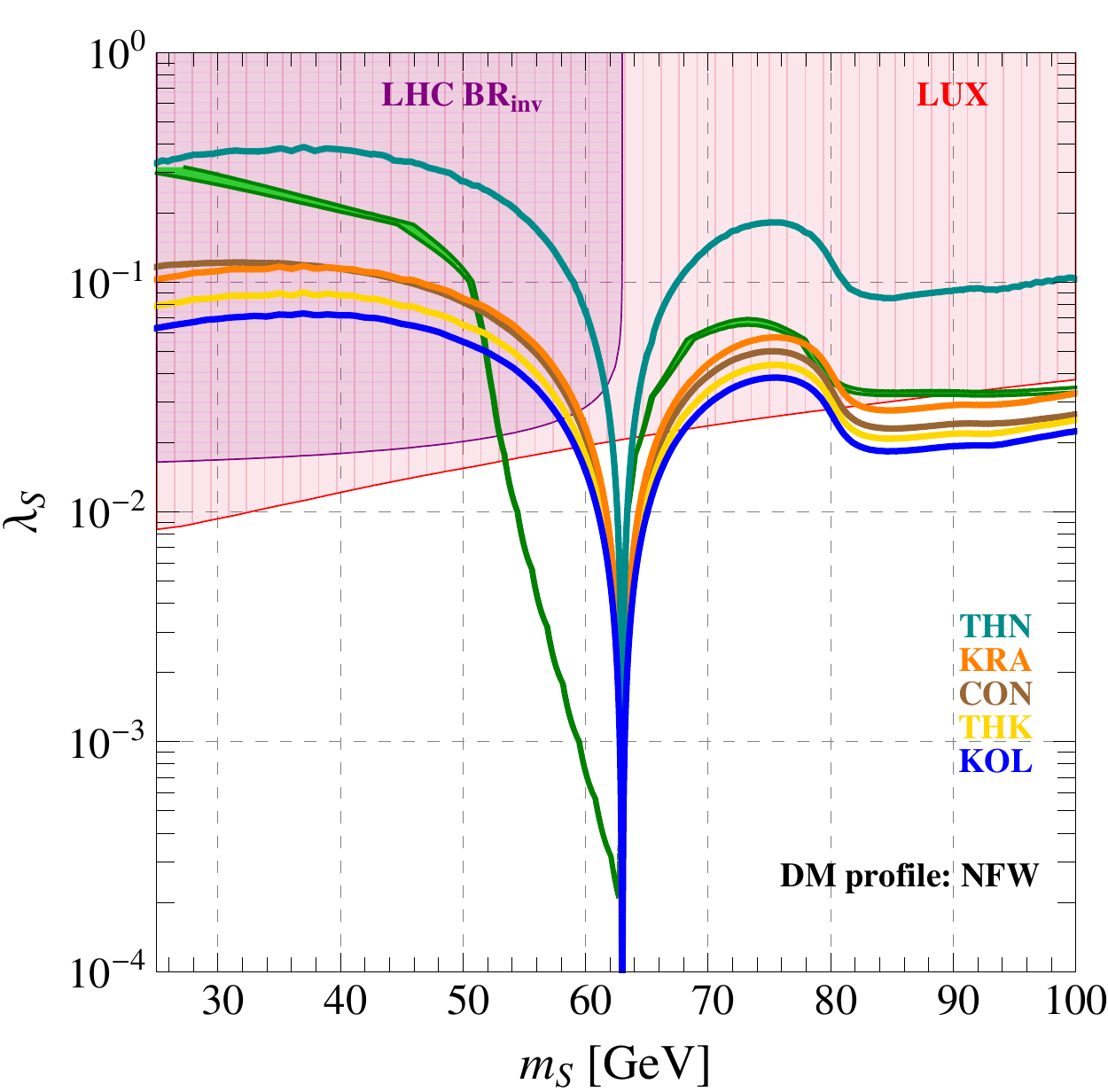}
\endminipage
\minipage{0.5\textwidth}
  \includegraphics[width=\linewidth]{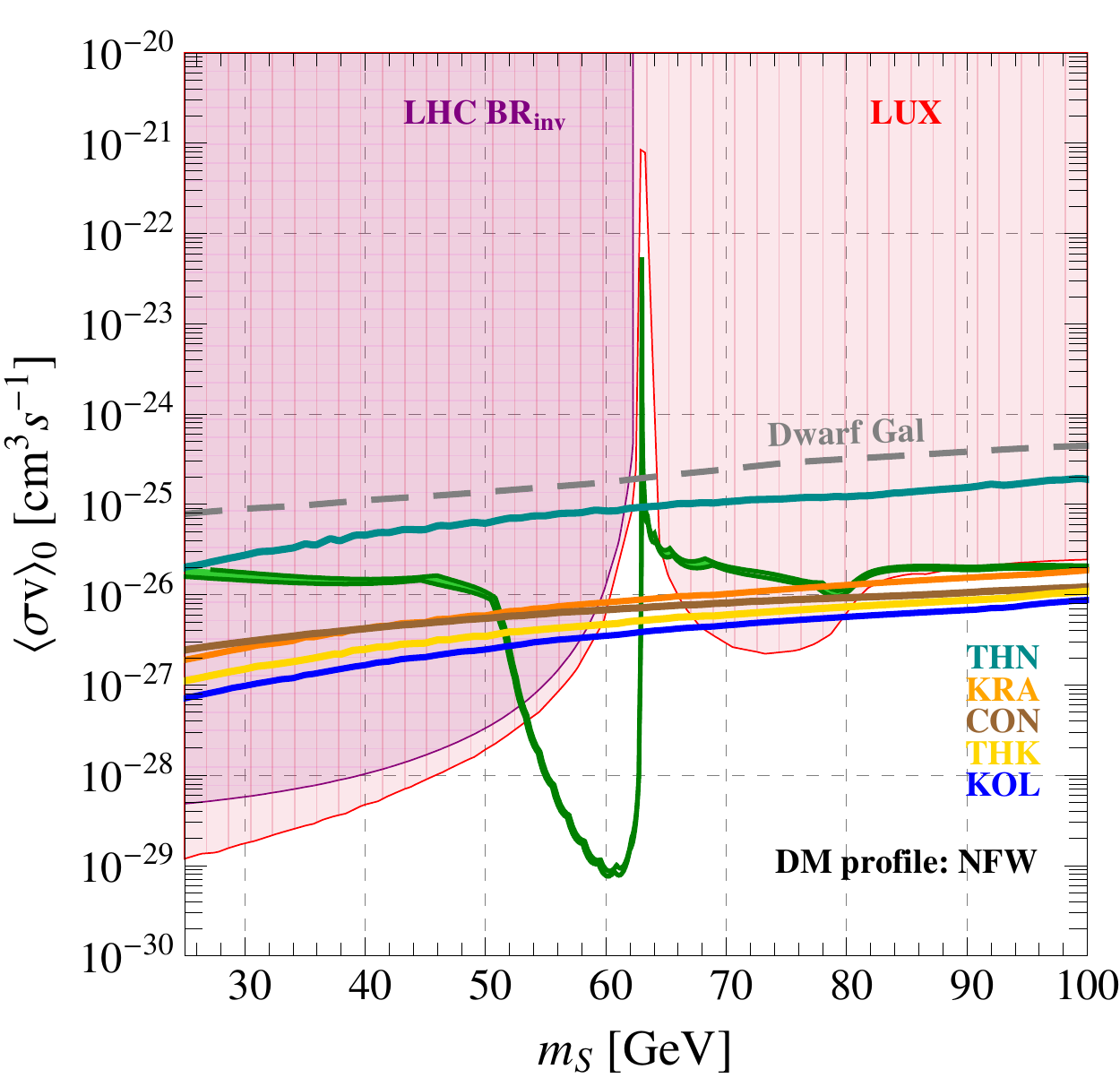}
\endminipage
 \caption{\textit{Bounds on the scalar Higgs portal model in the low-mass region.
 The green strip represents the $3\sigma$ 
 band reproducing the correct amount of relic density as measured by the Planck
 experiment. We show the region excluded at 95\% C.L. by the LHC considering the invisible branching ratio of the Higgs (purple region, horizontal meshes) and the region excluded at 95\% C.L.
 by the LUX experiment considering direct detection of DM (red region, vertical meshes).
 In addition, we show the constraints obtained from the antiproton data considering as a benchmark model for the DM density the NFW profile. Different propagation models are displayed with different colors [from bottom (more stringent) to top (less stringent): KOL (blue), THK (yellow), KRA (orange), CON (brown), THN (dark cyan); see text for details]. The region above each one of these curves is excluded. The dashed gray line represents the bound placed by the Fermi-LAT experiment using the gamma-ray data  from dwarf galaxies (see text).
 {\underline{Left panel}}. 
 Bounds on the parameter space $(m_{\rm S},\lambda_{\rm S})$.
 {\underline{Right panel}}.  Bounds on the thermally averaged annihilation cross-section times relative velocity at zero temperature.
 }}\label{fig:ScalarPortalLowMass}
\end{figure*}

\begin{figure*}[!htb]
\minipage{0.5\textwidth}
  \includegraphics[width=\linewidth]{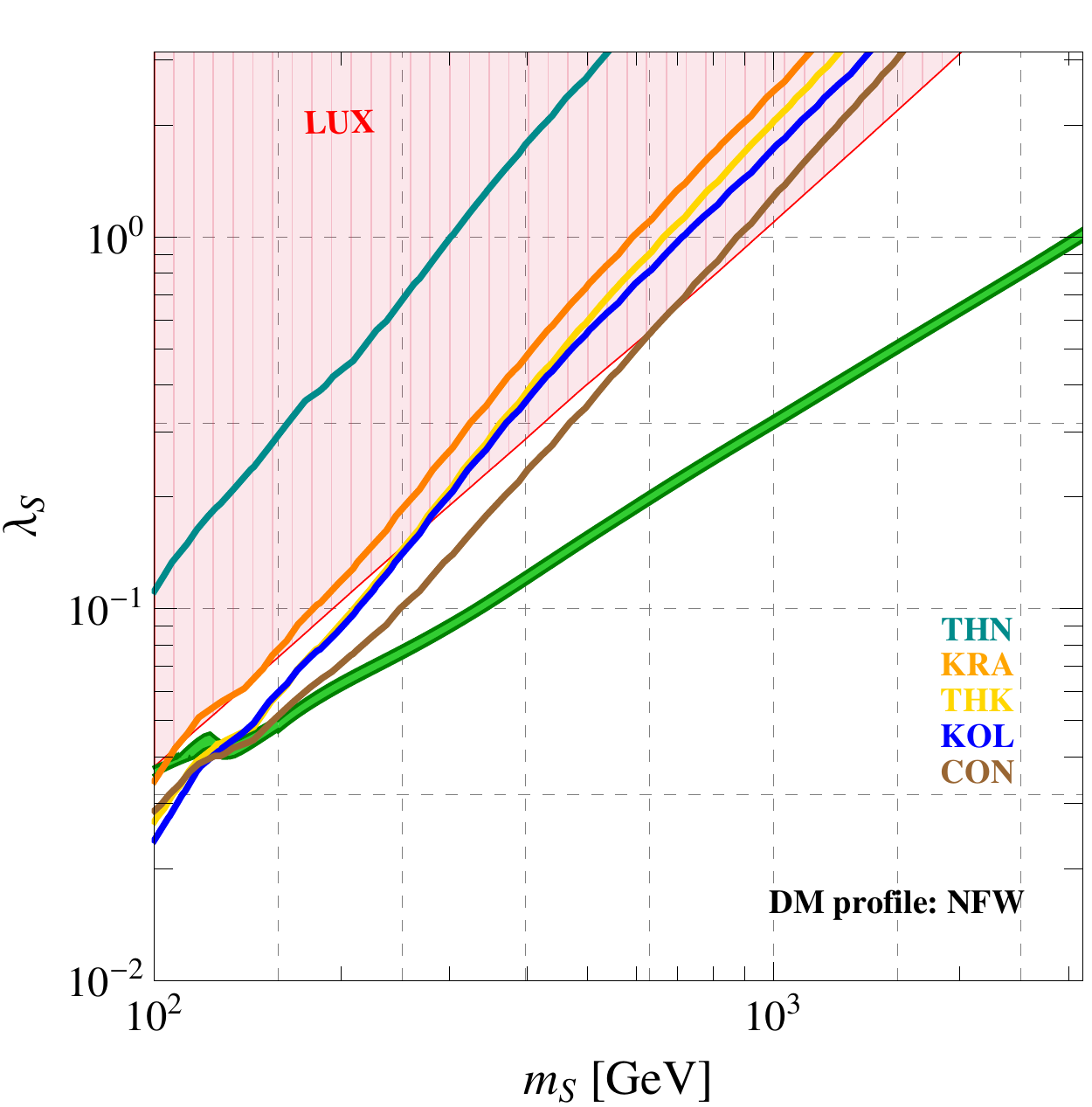}
\endminipage
\minipage{0.5\textwidth}
  \includegraphics[width=\linewidth]{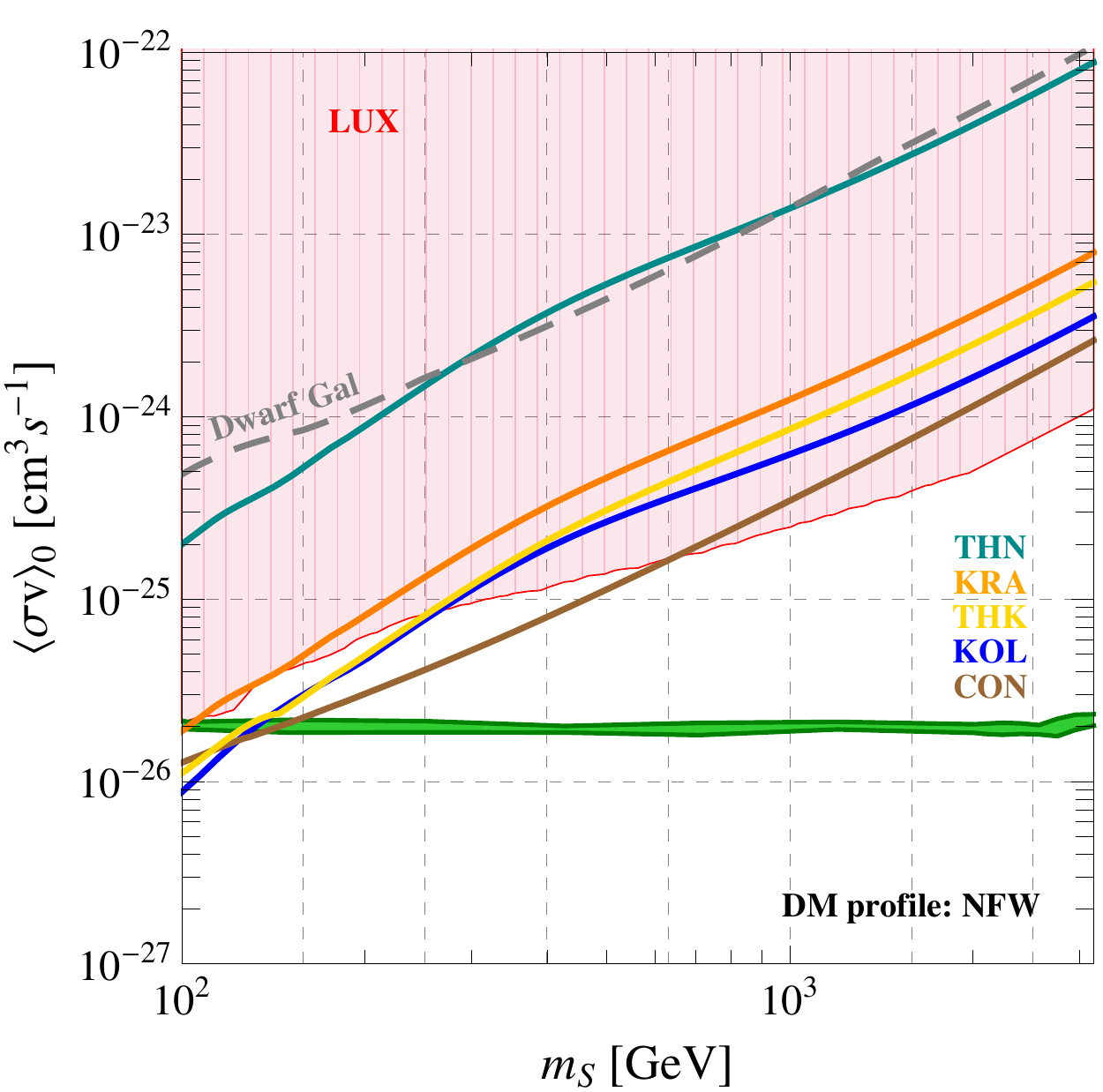}
\endminipage
 \caption{\textit{
 The same as in fig.~\ref{fig:ScalarPortalLowMass}, but for large values of the DM mass.
 }}\label{fig:ScalarPortalHighMass}
\end{figure*}

In this section, we study the antiproton bound on varying the propagation model, according to table~\ref{tab:models}. We show our results in fig.~\ref{fig:ScalarPortalLowMass} and 
in fig.~\ref{fig:ScalarPortalHighMass} where, for definiteness, we use the NFW
DM density profile \cite{Navarro:1995iw} (see fig.~\ref{fig:profiles}, eq.~(\ref{eq:NFW}), and table~\ref{tab:profiles} below). 

Let us start the discussion with some general comments.
The region that reproduces the correct value of relic density is represented by a green strip, while
the regions excluded by the LHC and LUX experiments are shaded, respectively, in purple and red.
In fig.~\ref{fig:ScalarPortalLowMass} we focus on small values for the DM mass, i.e. $m_{\rm S}\in [25, 100]$ GeV, in order to emphasize the role of the antiproton bound in the region close to the Higgs resonance. 
In the left panel of fig.~\ref{fig:ScalarPortalLowMass} we show our results in the parameter space $(m_{\rm S}, \lambda_{\rm S})$; considering  the green strip that reproduces the correct value of relic abundance, 
the resonant region is immediately recognizable because of the usual funnel-shaped form.
As already discussed in section~\ref{sec:RelicDensity}, this  region extends also for values of the DM mass smaller than $m_{h}/2 = 63$ GeV as a consequence of thermal effects during the freeze-out epoch,
and this feature clearly emerges in the plot from the result of our numerical analysis; moreover, notice that both the bound on the invisible Higgs decay width and the spin-independent DM-nucleon cross-section can not rule out this region because of the kinematical reasons discussed in sections~\ref{sec:LHCbound},~\ref{sec:DD}.  In the right panel of fig.~\ref{fig:ScalarPortalLowMass}, we translate our analysis in the plane $(m_{\rm S}, \langle \sigma v_{\rm rel}\rangle_0)$. Away from the resonance the value of  $\langle \sigma v_{\rm rel}\rangle_0$ that reproduces the observed relic abundance is close to the usual WIMP-miracle cross-section 
$\langle \sigma v_{\rm rel}\rangle_0 \approx 2\times 10^{-26}$ cm$^3$s$^{-1}$. Close to the resonance, on the contrary, this value is distorted by the presence of the aforementioned thermal effects that move the position of the resonance during the freeze-out epoch. In particular, we find that
for $50 \lesssim m_{\rm S} \lesssim m_h/2$ the thermally averaged annihilation cross-section times relative velocity today can be as small as $\langle \sigma v_{\rm rel}\rangle_0 \approx 10^{-29}$ cm$^3$s$^{-1}$, while for $m_{\rm S}\approx m_h/2$ we have 
$\langle \sigma v_{\rm rel}\rangle_0 \approx 10^{-22}$ cm$^3$s$^{-1}$.
In fig.~\ref{fig:ScalarPortalHighMass} we focus on large values for the DM mass, i.e. $m_{\rm S}\in [100, 3000]$ GeV. As in fig.~\ref{fig:ScalarPortalLowMass}, we show the constraints placed by our phenomenological analysis in the plane 
$(m_{\rm S},\lambda_{\rm S})$, left panel, and $(m_{\rm S},\langle \sigma v_{\rm rel}\rangle_0)$, right panel.

In fig.~\ref{fig:ScalarPortalLowMass} and fig.~\ref{fig:ScalarPortalHighMass} we show the antiproton bound that corresponds to the five propagation models defined in section~\ref{sec:Analysis}, and we use the same color code introduced in table~\ref{tab:models}. 
The results of our analysis point towards the following remarks.
\begin{itemize}

\item  On a general ground, we find that the KOL, THK, CON and KRA propagation models give similar bound, while the THN propagation model places the weaker constraint. In greater detail, in the low mass region the KOL model, characterized by a strong re-acceleration, gives the strongest constraint. On the contrary in the high mass region the CON model, characterized by a strong convective wind, provides the most stringent bound. The presence of strong convective effects, in fact, hardens the antiproton flux thus 
  leading to stronger constraints on heavier DM models \cite{Evoli:2011id}.
It is worth noticing that  the THN model, based on a  thin diffusion zone, 
is  disfavored by recent studies on synchrotron emission, radio
maps and low energy positron spectrum \cite{DiBernardo:2012zu}.

\item In the low mass region the antiproton bound is competitive with 
the bound obtained from direct detection and invisible Higgs decay width. In particular, we find that the
bound from antiproton is the only one able to rule out the resonant region with $m_{\rm S}\approx m_h/2$.
Let us stress once again that this specific value for the DM mass would be otherwise inaccessible.
On the one hand, in fact, the invisible Higgs branching ratio goes to zero moving towards the kinematical 
threshold $m_{\rm S}= m_h/2$; on the other one, the square of the momentum transferred
in a typical DM-nucleus elastic scattering always satisfies the condition $-q^2 \ll m_h^2$, with 
$q^2 = -2m_{\rm Xe}E_{\rm rec}$ where the mass of a Xenon nucleus is $m_{\rm Xe}= 121$ GeV and for the typical recoil energy one has $E_{\rm rec} \sim $ few keV.

\item In the high mass region the antiproton bound obtained using the KOL, THK, CON and KRA propagation models
is competitive with the exclusion curve traced by the LUX experiment. In particular, as clear from 
the right panel of fig.~\ref{fig:ScalarPortalHighMass}, using the KOL, THK and CON 
propagation models it is possible to probe the thermal cross-section up to $m_{\rm S}\approx 160$ GeV.

\item For comparison, we show in the right panel of figs.~\ref{fig:ScalarPortalLowMass},~\ref{fig:ScalarPortalHighMass} the 95\% C.L. exclusion curve obtained considering the measurement of the gamma-ray flux from the dwarf spheroidal satellite galaxies of the Milky Way \cite{Ackermann:2013yva}.
These dwarf galaxies are some of the most DM-dominated objects known, and -- because of their proximity, high DM content, and lack of astrophysical backgrounds -- they are usually considered to be the most promising targets for the indirect detection of DM via gamma rays. For simplicity, in our analysis we used only the data from  the Draco dwarf spheroidal galaxy since it gives the strongest constraint. Let us now describe in more detail our approach. In order to use the result of ref.~\cite{Ackermann:2013yva}, first we computed the gamma-ray flux from the Draco  dwarf spheroidal galaxy
in the Higgs portal model under scrutiny, combining all the different annihilation channels including three-body final states. Then, for each value of the DM mass, 
we compared the gamma-ray flux previously obtained with the 95\% C.L.
exclusion limit in each of the 24 energy bins analyzed in fig.~2 of ref.~\cite{Ackermann:2013yva}.
Finally, we extracted the bound on 
the cross-section from the energy bin that provides the strongest constraint. We find that, both in the low- and high-mass regions, the bound from antiproton that we obtain using the  KOL, THK, CON and KRA propagation models is more than one order of magnitude stronger than the bound  obtained from the analysis of the gamma-ray flux measured from the Draco dwarf spheroidal galaxy. Needless to say, a more detailed analysis would  
require to include all the 25 dwarf spheroidal galaxies studied in ref.~\cite{Ackermann:2013yva} together with a more careful investigation of the systematic errors involved. This task goes well beyond the purpose of the simple estimation that we derived in this work, and will be left for future investigation.

\end{itemize}

In conclusion, we have found that the antiproton bound provides a strong constraint on the parameter space of the scalar Higgs portal model introduced in section~\ref{sec:HiggsPortal}. Remarkably, the constraining power of the antiproton data is comparable to the 
exclusion curves placed by the LHC and LUX experiments in particular for $m_{\rm S}\gtrsim 50$ GeV. 
Most importantly, the antiproton bound is the only one able to rule out the Higgs resonant region for $m_{\rm S}\approx 63$ GeV. This conclusion does not strongly depend on the model used to describe the dynamics underlying the propagation of  
charged particle in the Galaxy; in particular, we have shown that the KOL, THK, CON and KRA propagation setups 
give, in magnitude, similar bounds.
\begin{figure*}[!htb]
\minipage{0.5\textwidth}
  \includegraphics[width=\linewidth]{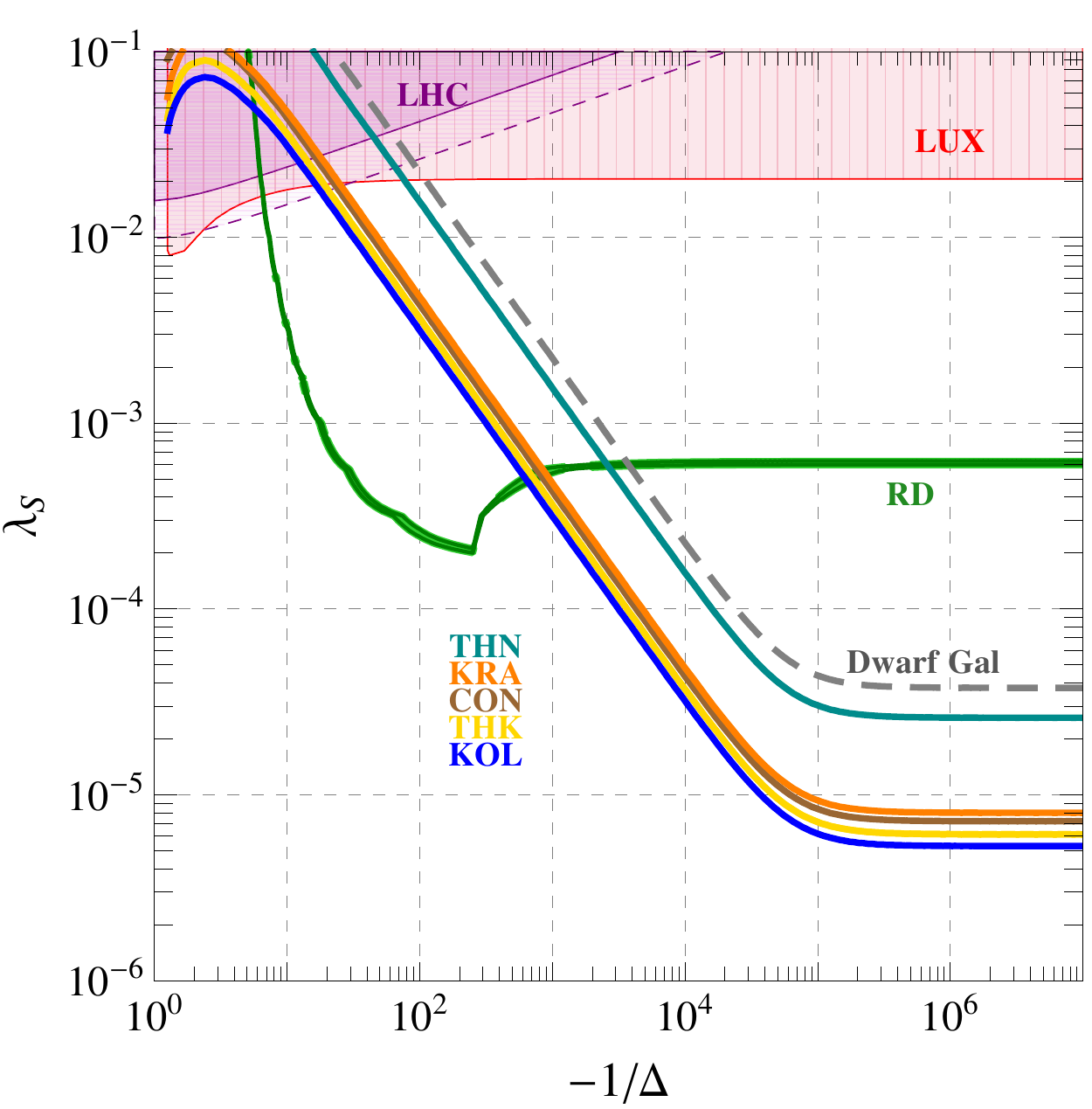}
\endminipage
\minipage{0.5\textwidth}\hspace{-0.2 cm}
  \includegraphics[width=\linewidth]{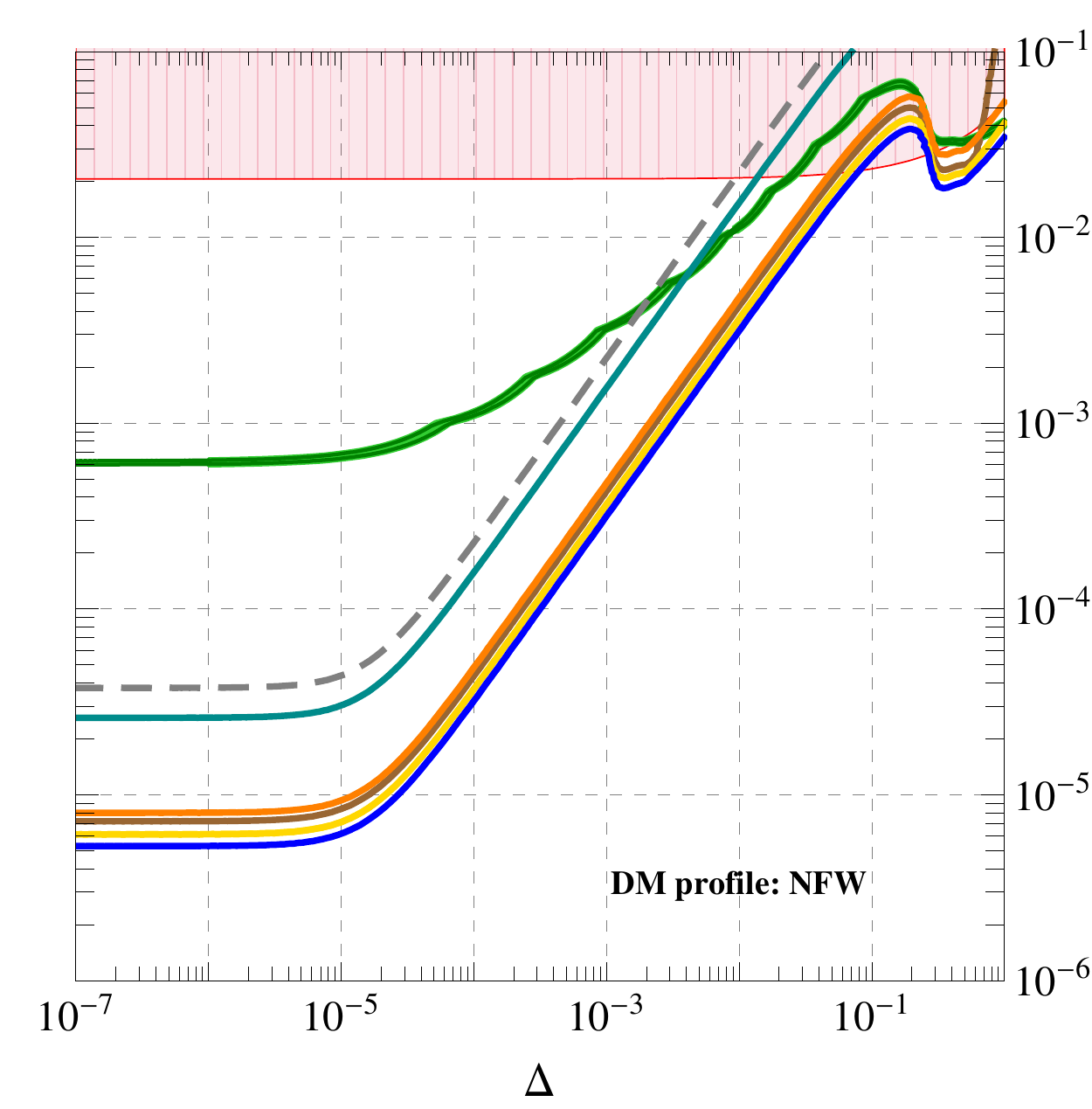}
\endminipage
 \caption{\textit{
 The same as in fig.~\ref{fig:ScalarPortalLowMass}, but with a special focus on the Higgs resonance.
 }}\label{fig:ResZoom}
\end{figure*} 
In order to stress this point we show in fig.~\ref{fig:ResZoom}, following ref.~\cite{Feng:2014vea}, a zoom on the Higgs resonant region. 
We introduce the variable $\Delta \equiv (2 m_{\rm S} - m_h)/m_h$, and we present our constraints in the plane $(\Delta, \lambda_{\rm S})$. 
From this point of view, it is clear that the bound we get from the antiproton data is by far the most stringent if compared with LHC and LUX results. 
The only region left unconstrained by the antiproton bound is the small mass window with $10 \lesssim -1/\Delta\lesssim 10^{3}$
which corresponds to $56.7 \lesssim m_{\rm S}\lesssim 62.9$ GeV with $10^{-4} \lesssim \lambda_{\rm S}\lesssim 10^{-2}$.
In this region the 
position of the Higgs resonance is subject to thermal effects; as previously discussed, in this small mass window 
the resonant annihilation cross-section reproducing in the early Universe 
the correct relic abundance corresponds to an off-resonant value in today's annihilations in the Galactic halo.

In this section we extracted the antiproton bound using the standard NFW profile in order to describe 
 the density distribution of DM in the Galaxy.
In the next section, we will discuss the impact of different DM halo profiles.

\subsection{On the impact of different DM density profiles}\label{sec:diffProf}

In this section we explore the impact of different DM density profile on the analysis of the antiproton data.
In addition to the NFW profile \cite{Navarro:1995iw} already used in section~\ref{sec:diffProp}, we repeat our analysis using the Einasto \cite{Navarro:2008kc,Graham:2005xx} and the Isothermal profile \cite{Burkert:1995yz}.
The former -- similar to the NFW profile and  characterized by a DM density distribution peaked towards the Galactic center -- is favored by the latest standard numerical simulations \cite{Navarro:2003ew,Graham:2006ae} while the latter -- characterized by a constant core -- seems to be in agreement with the numerical simulations that include baryons, because of large exchange of angular momentum between the gas and DM particles \cite{ElZant:2003rp}.
We show these three DM density distributions in fig.~\ref{fig:profiles} and eqs.~(\ref{eq:NFW},~\ref{eq:Ein},~\ref{eq:Iso}), while in table~\ref{tab:profiles}
we collect the numerical values of the parameters that enter in their definitions.

  \begin{minipage}{\textwidth}
  \begin{minipage}[b]{0.45\textwidth}
    \centering
   \includegraphics[width= \linewidth]{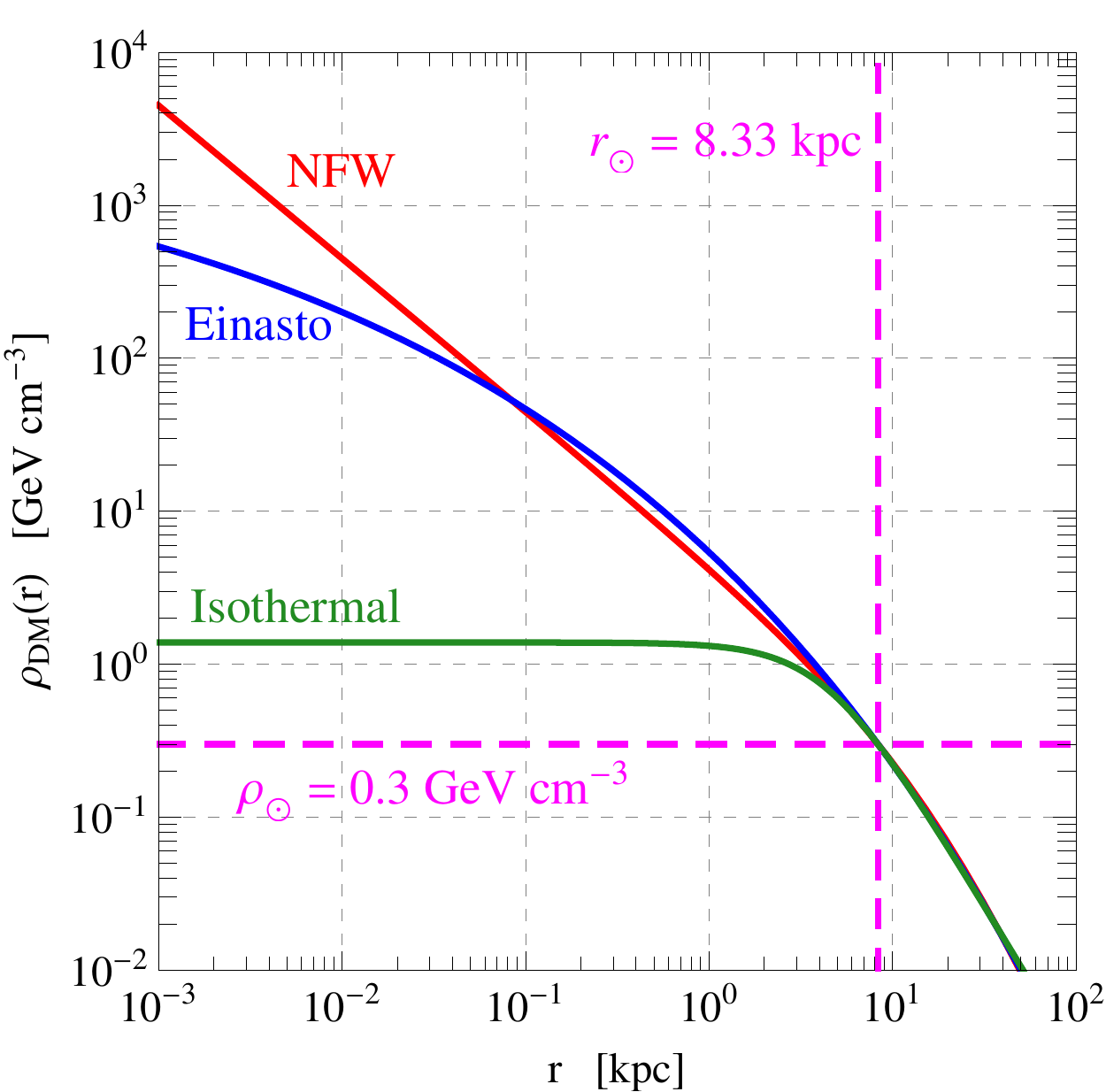}
    \captionof{figure}{
    \textit{DM density distributions in eqs.~(\ref{eq:NFW},~\ref{eq:Ein},~\ref{eq:Iso}).}}\label{fig:profiles}
  \end{minipage}
  \hspace{0.01 cm}
  \begin{minipage}[b]{0.45\textwidth}
    \centering
\begin{eqnarray}
\rho_{\rm NFW}(r) &=&  \rho_s \frac{r_s}{r}\left(1 + \frac{r}{r_s}\right)^{-2}~, \label{eq:NFW} \\
\rho_{\rm Ein}(r) &=&   \rho_s\,e^{\left\{-\frac{2}{\alpha}\left[\left(\frac{r}{r_s}\right)^{\alpha} - 1\right] \right\}}~,\label{eq:Ein} \\
\rho_{\rm Iso}(r) &=&   \frac{\rho_s}{1+\left(\frac{r}{r_s}\right)^2}\label{eq:Iso}~.
\end{eqnarray}
\begin{tabular}{|c|c|c|}\hline
     \textbf{DM halo}   & $r_s$ [kpc] & $\rho_s$ [GeVcm$^{-3}$]  \\ \hline
      {\color{red}{NFW}}             & 24.42 & 0.184 \\ \hline
     {\color{blue}{Ein}}                & 28.44 & 0.033  \\ \hline
      {\color{forestgreen}{Iso}}    &   4.38 &  1.385  \\ \hline
      \end{tabular}
 \captionof{table}{\textit{Parameters defining the DM density distributions in fig.~\ref{fig:profiles}.
 For the Einasto profile, $\alpha = 0.17$.}}\label{tab:profiles}
\vspace{0.01 cm}
    \end{minipage}
  \end{minipage}
  \vspace{0.2 cm}
  
\begin{figure}[!htb!]
\vspace{-0.8cm}
\begin{center}
\hspace*{-0.65cm} 
\begin{minipage}{0.5\linewidth}
\begin{center}
	\includegraphics[width=\linewidth]{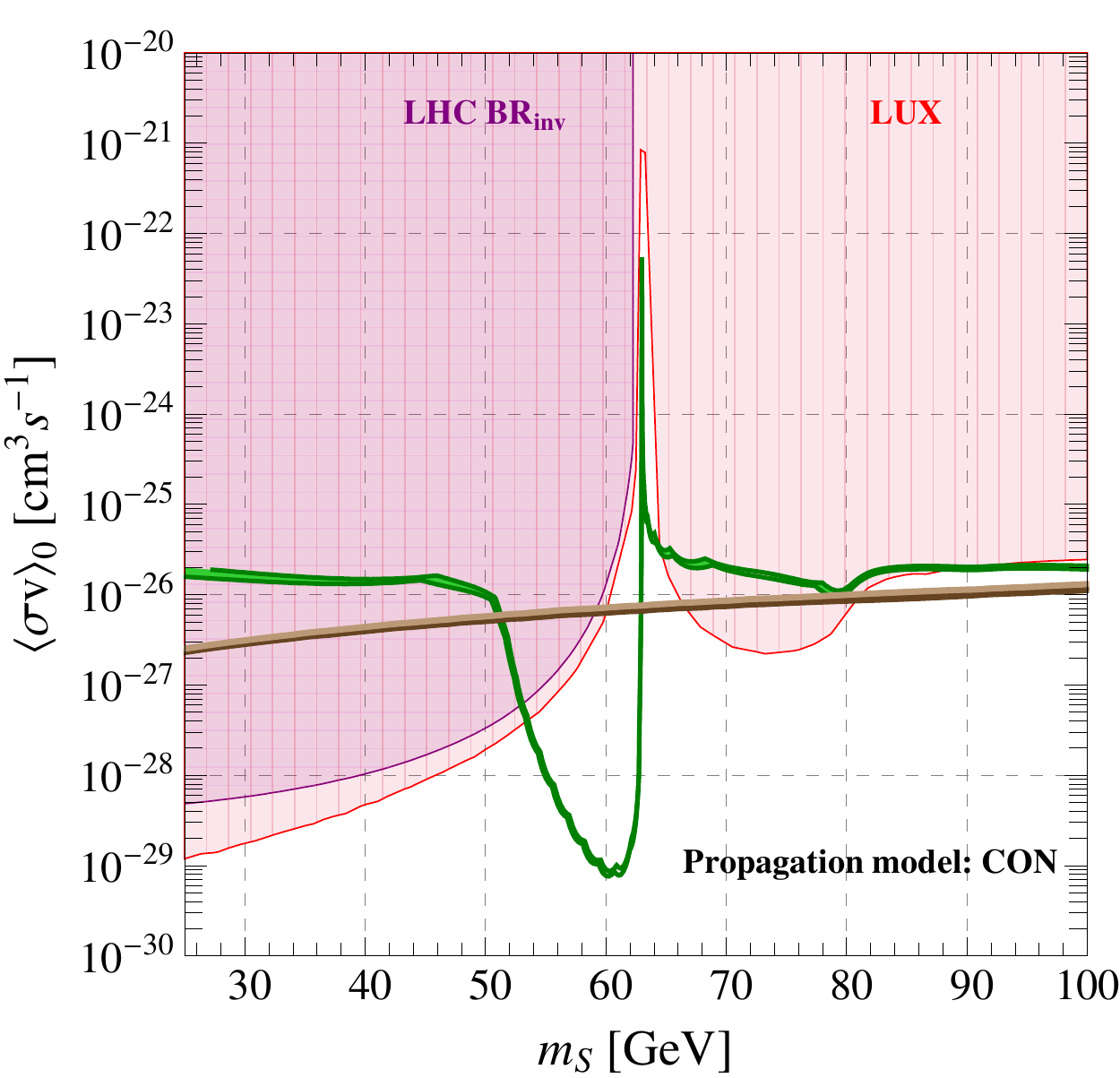}
\end{center}
\end{minipage}
\begin{minipage}{0.5\linewidth}
\begin{center}
	\includegraphics[width=\linewidth]{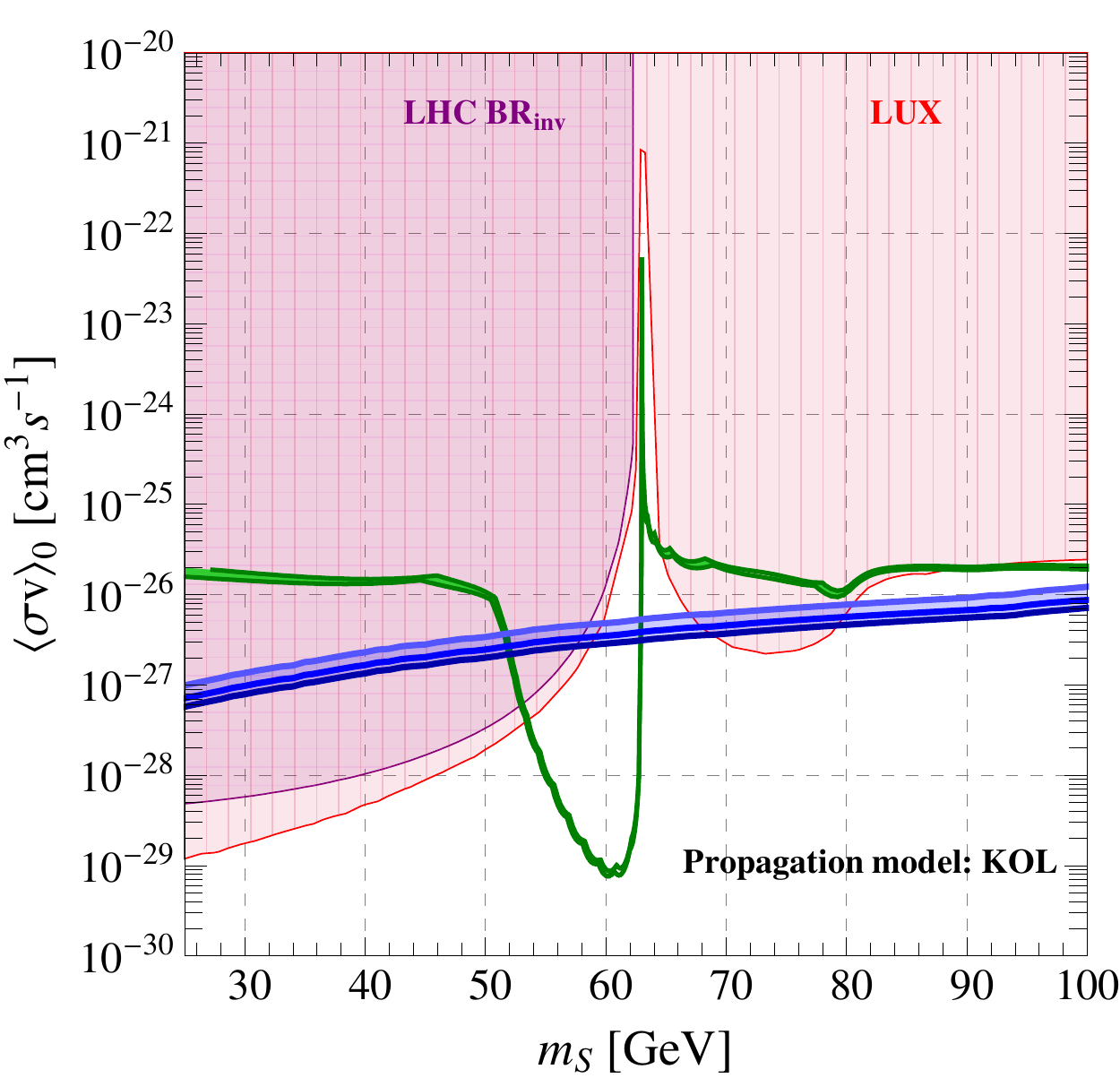}
\end{center}
\end{minipage}
\\[0.1cm]
\hspace*{-0.65cm} 
\begin{minipage}{0.5\linewidth}
\begin{center}
	\includegraphics[width=\linewidth]{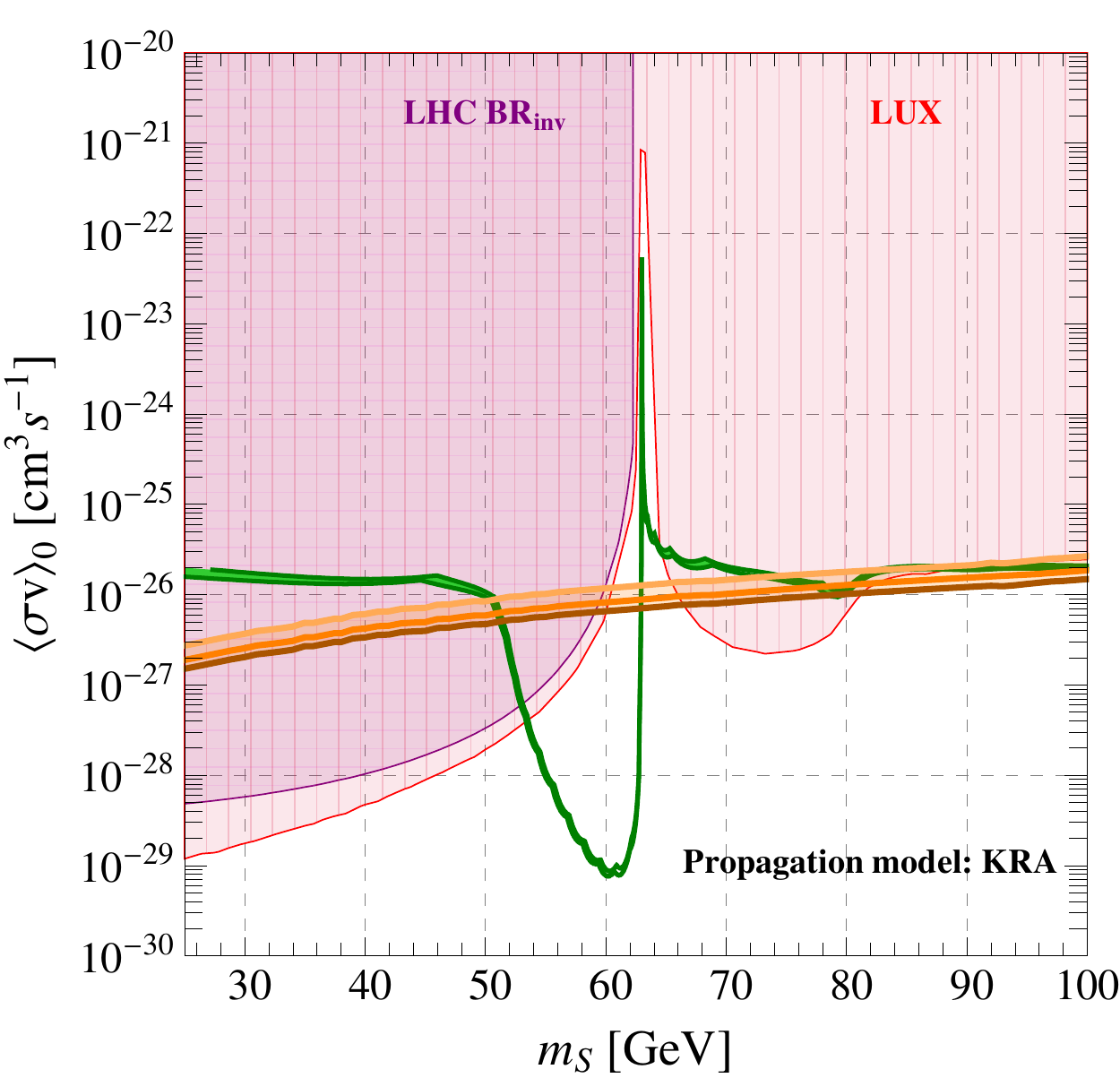}
\end{center}
\end{minipage}
\begin{minipage}{0.5\linewidth}
\begin{center}
	\includegraphics[width=\linewidth]{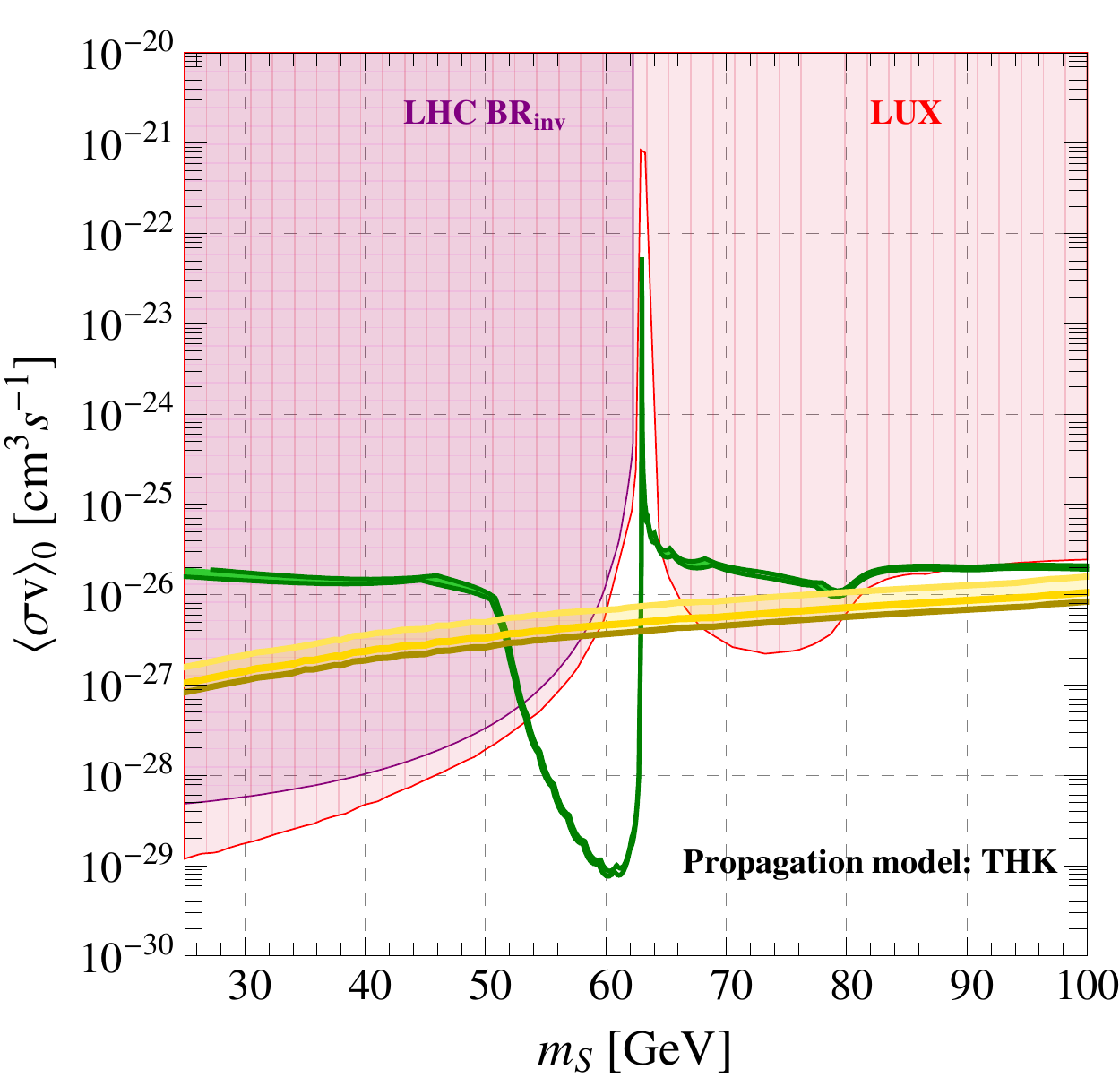}
\end{center}
\end{minipage}
\end{center}
\vspace*{-0.2cm}
\caption{
\textit{
Antiproton bounds on the thermally averaged annihilation cross-section times relative velocity for different DM density distributions, namely the Einasto, NFW and Isothermal profiles.  The most (less) stringent bound corresponds to the Einasto (Isothermal) profile. For comparison, we show the impact of different profiles for the four propagation models CON, KOL, THK, KRA (from top left, clockwise). The color code -- as well as the other bounds from the LHC and LUX experiments -- follows fig.~\ref{fig:ScalarPortalLowMass}. The bound obtained for the THN propagation model does not depend on the DM density distribution, see text for details. 
}}\label{fig:CombinedPlotLowMass} 
\end{figure}

\begin{figure}[!htb!]
\vspace{-0.8cm}
\begin{center}
\hspace*{-0.65cm} 
\begin{minipage}{0.5\linewidth}
\begin{center}
	\includegraphics[width=\linewidth]{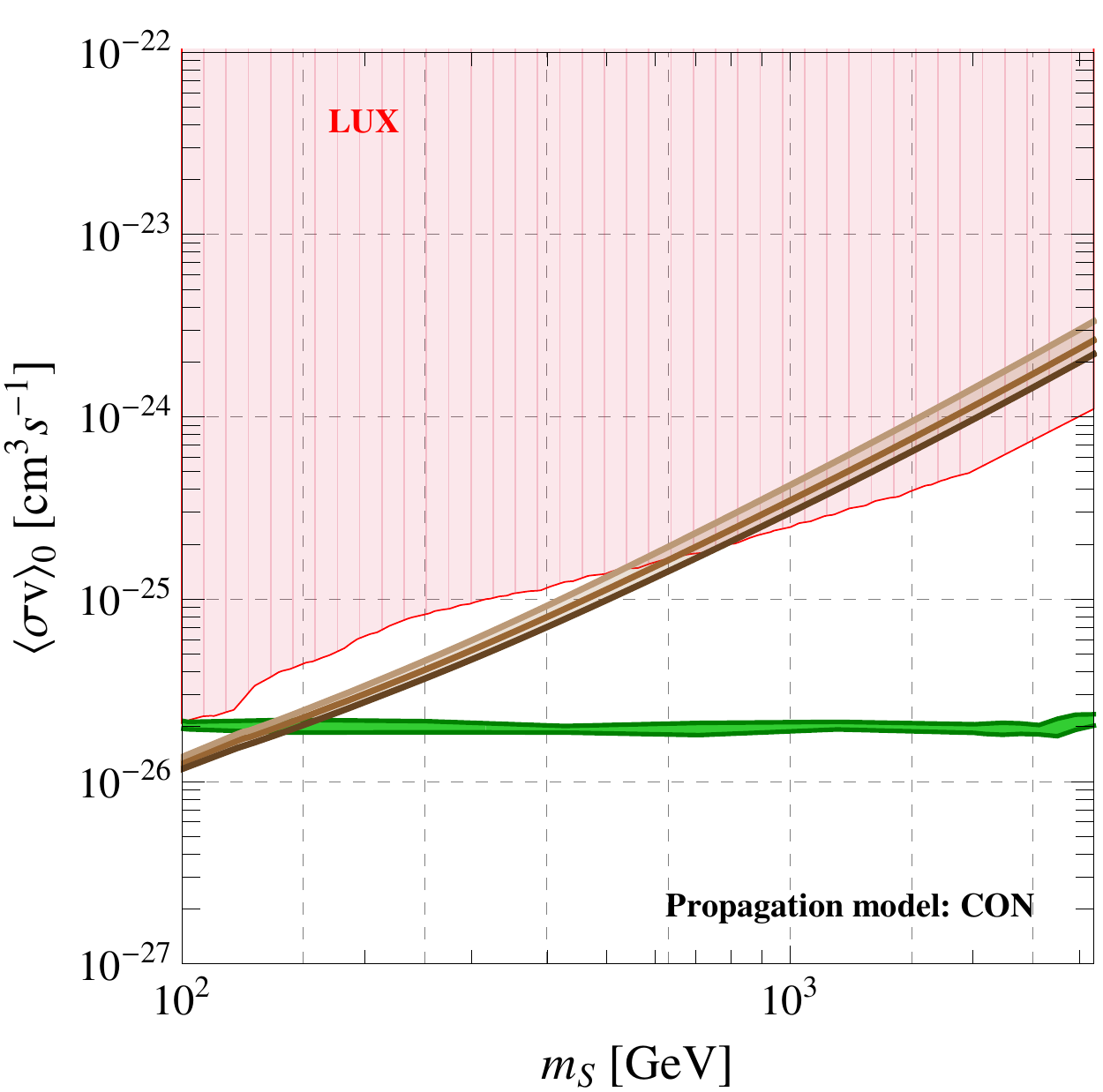}
\end{center}
\end{minipage}
\begin{minipage}{0.5\linewidth}
\begin{center}
	\includegraphics[width=\linewidth]{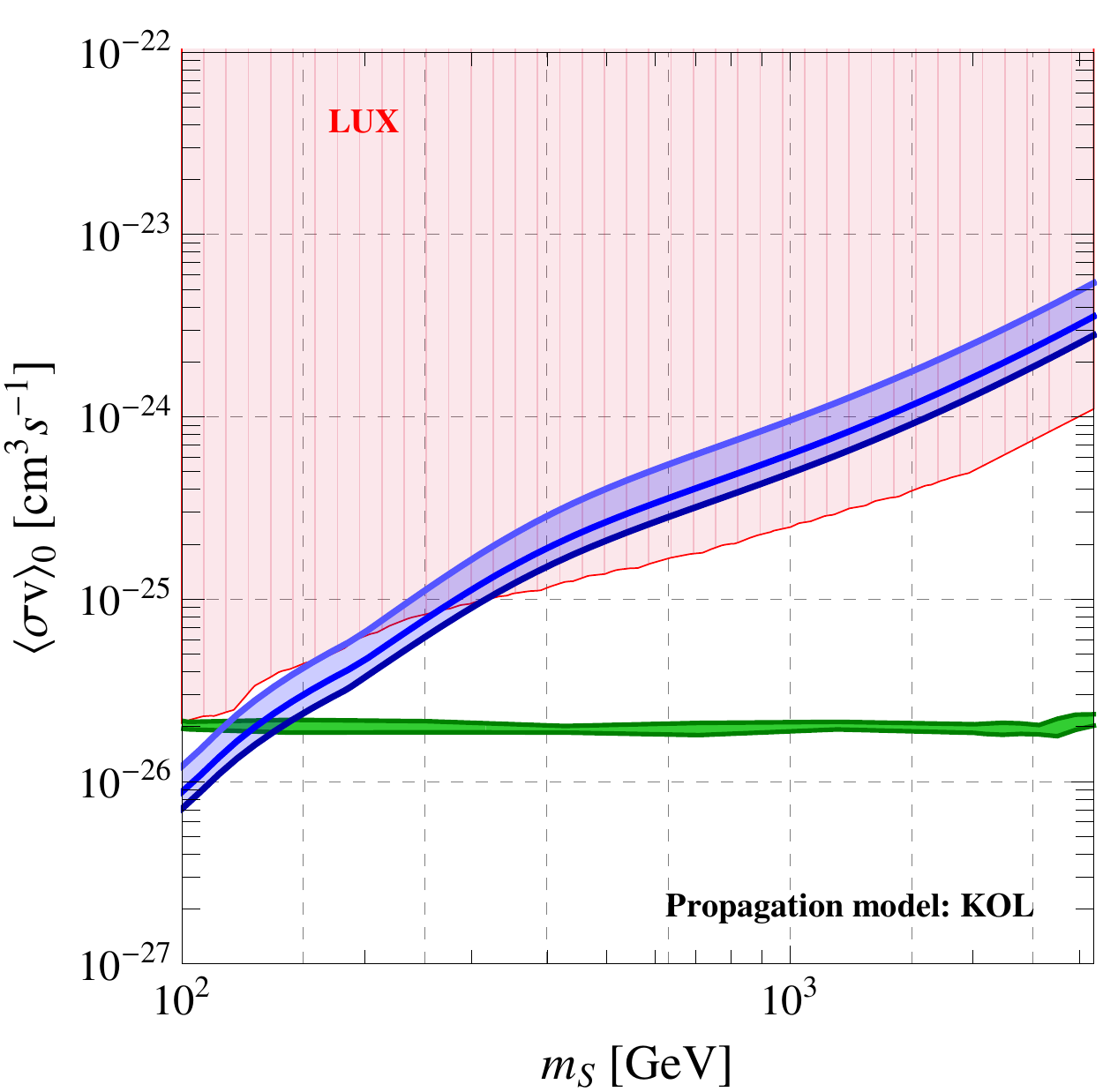}
\end{center}
\end{minipage}
\\[0.1cm]
\hspace*{-0.65cm} 
\begin{minipage}{0.5\linewidth}
\begin{center}
	\includegraphics[width=\linewidth]{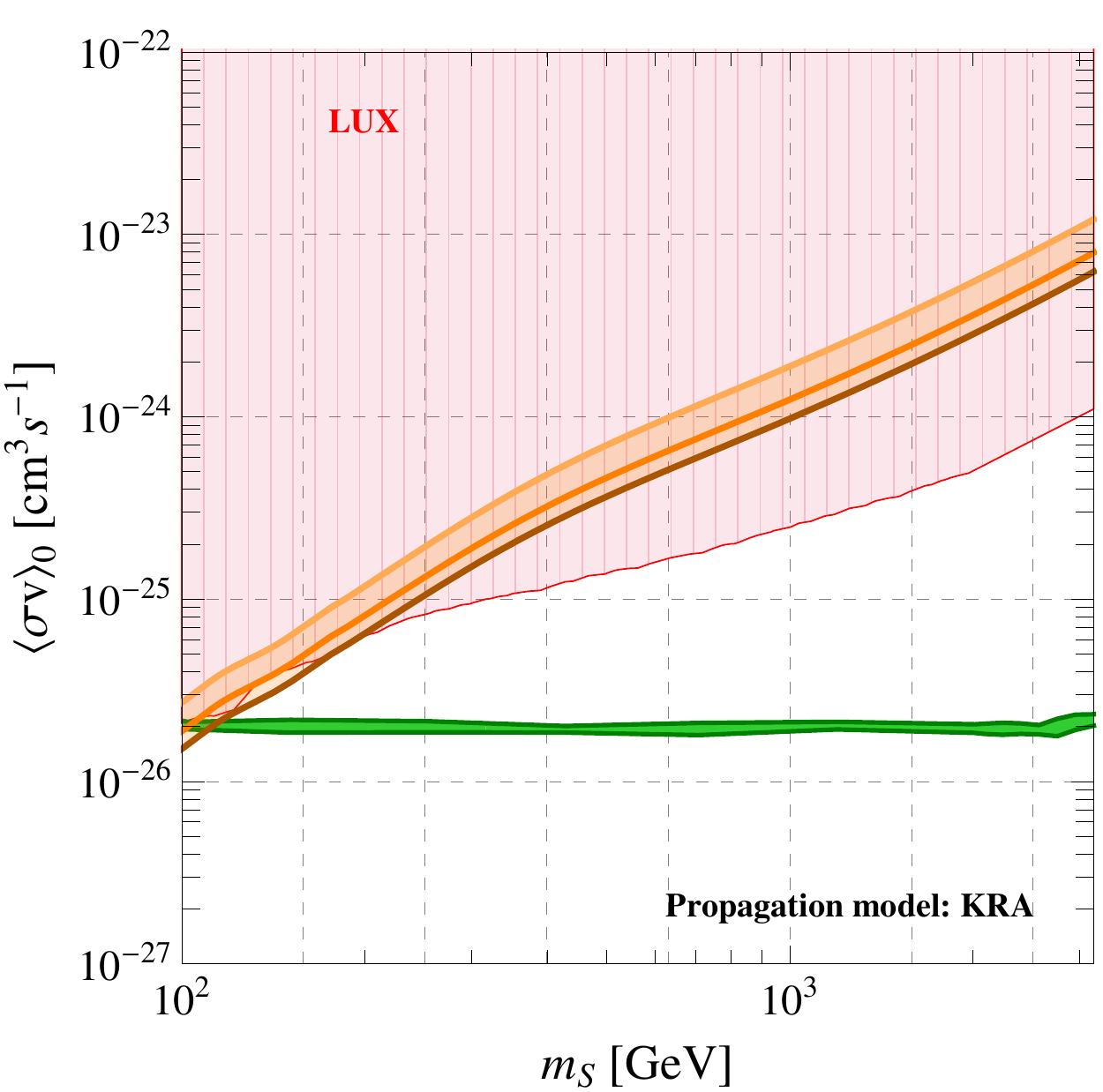}
\end{center}
\end{minipage}
\begin{minipage}{0.5\linewidth}
\begin{center}
	\includegraphics[width=\linewidth]{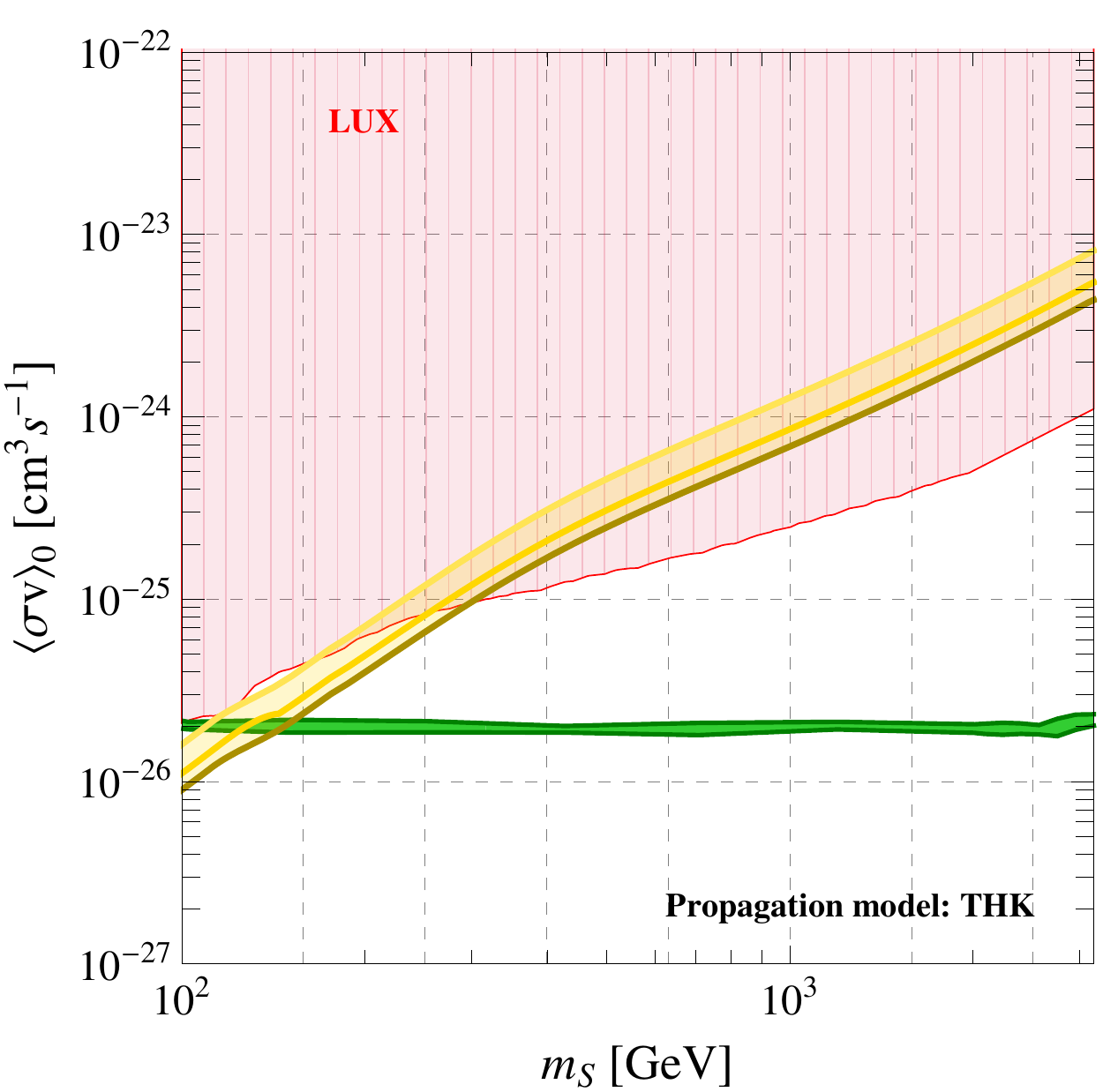}
\end{center}
\end{minipage}
\end{center}
\vspace*{-0.2cm}
\caption{
\textit{
 The same as in fig.~\ref{fig:CombinedPlotLowMass}, but for large values of the DM mass.
}}\label{fig:CombinedPlotHighMass} 
\end{figure}

We show our results in fig.~\ref{fig:CombinedPlotLowMass}, for the low mass region, and in fig.~\ref{fig:CombinedPlotHighMass}, for the high mass region. 
In both cases we focus on the plane $(m_{\rm S}, \langle \sigma v_{\rm rel}\rangle)$.

\begin{figure*}[!htb]
\minipage{0.5\textwidth}
  \includegraphics[width=\linewidth]{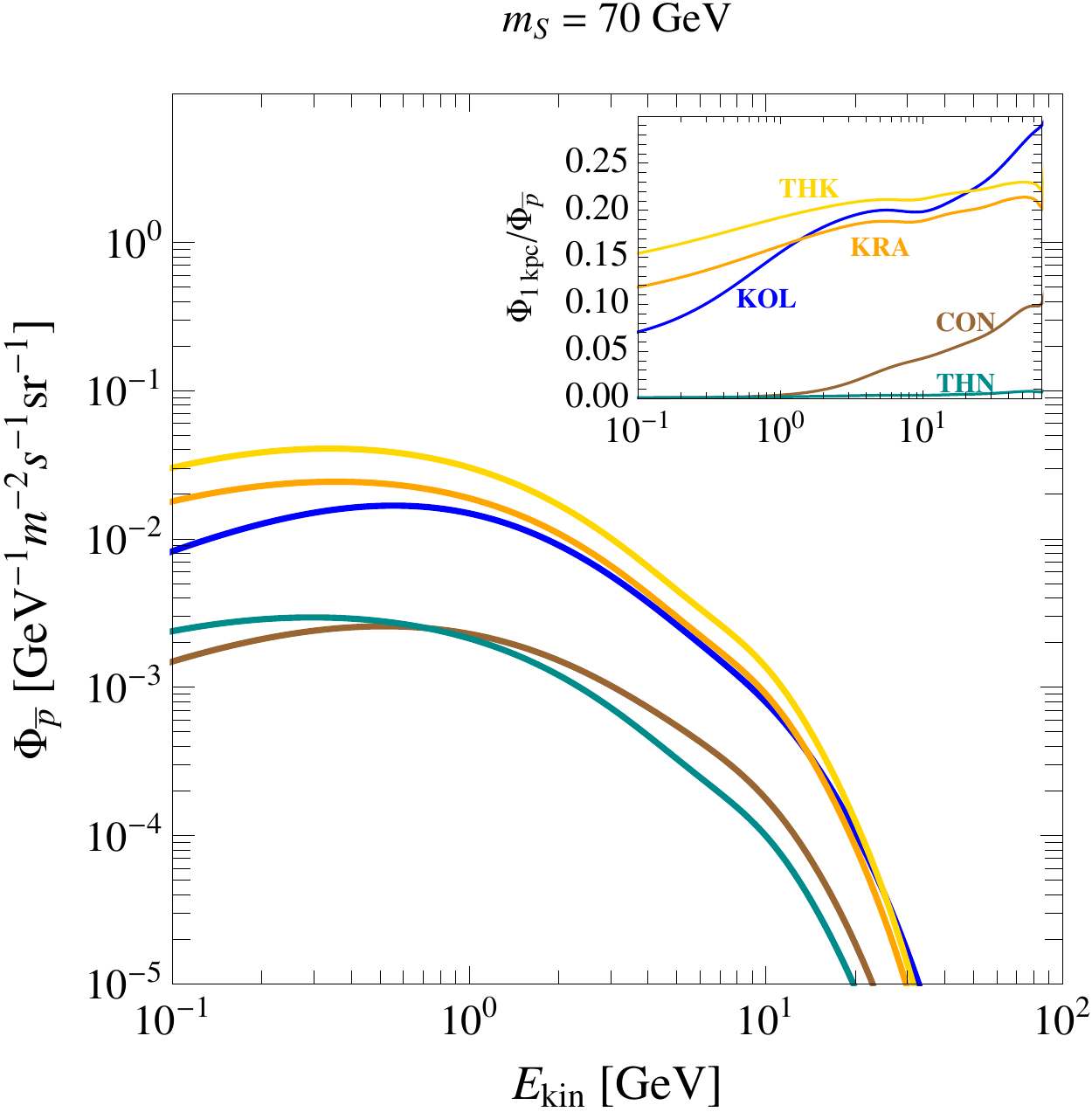}
\endminipage
\minipage{0.5\textwidth}
  \includegraphics[width=\linewidth]{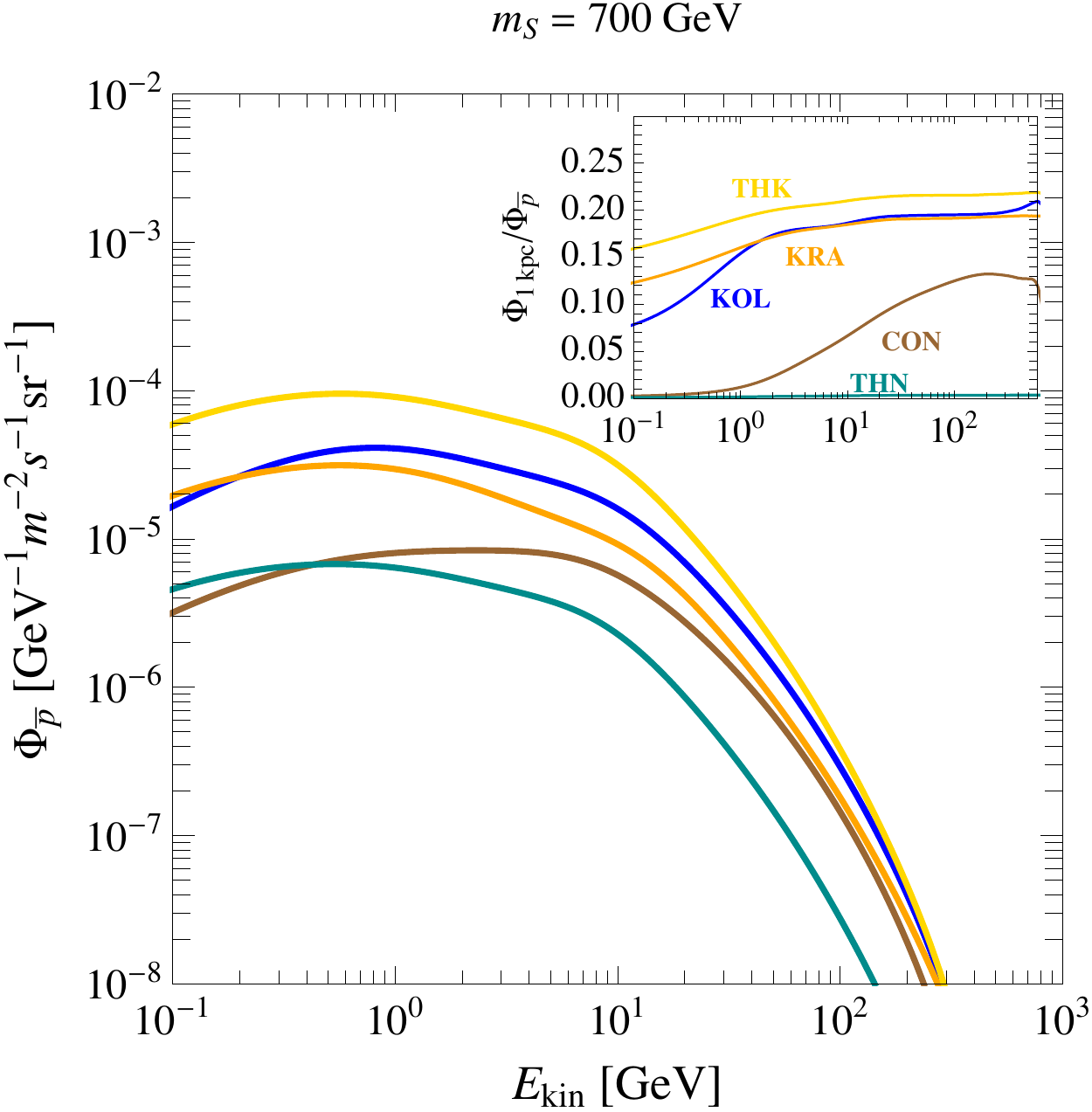}
\endminipage
 \caption{\textit{
 Local antiproton flux from DM annihilation for two representative values of DM mass and the five propagation models defined in  table~\ref{tab:models}.
 In the inset plots, we show the relative contribution to the total local flux 
 coming from a region enclosed within $1$ kpc from the Galactic center. We use the NFW profile, and we do not include solar modulation effects, $\Phi =0$ GV.
 }}\label{fig:RadialDependence}
\end{figure*}

Let us start our discussion pointing out an important argument to keep in mind for the rest of the section.
As a rule of thumb, one would expect that the antiproton flux from DM annihilation is larger for 
profiles  in which the DM density is enhanced towards the Galactic center while is smaller for density
distribution described by an isothermal sphere; as a consequence, one would na\"{\i}vely  guess that a common feature 
of the analysis is that the antiproton bound is always more (less) stringent for the NFW (Isothermal) profile.
In general, however, this conclusion turns out to be partially incorrect. What really matters, in fact, 
is  not the value of DM density at the Galactic center but
at the position where the antiprotons -- whose flux is measured on Earth -- are generated.
As one can easily imagine, further insights on this issue are strongly linked to the assumed propagation model,
and our analysis points towards the following results.
\begin{itemize}

\item We find that the antiproton bound obtained  using the THN model, based on a very thin diffusion zone,
does not significantly depend on the DM density profile, and we do not show the corresponding plot in figs.~\ref{fig:CombinedPlotLowMass},~\ref{fig:CombinedPlotHighMass}. In the THN model, in fact,
the antiproton flux from DM annihilation is dominated by local contributions where the three profiles are equivalent \cite{Evoli:2011id}.

\item As far as the antiproton bound obtained assuming the CON propagation model is concerned, 
we find that also in this case the impact of different DM density distribution is moderately negligible, as shown in the upper-left panel of figs.~\ref{fig:CombinedPlotLowMass},~\ref{fig:CombinedPlotHighMass}. As already observed in ref.~\cite{Evoli:2011id}, therefore, we argue that the uncertainty related to the DM distribution towards the Galactic
center has a negligible effect in the CON model in which the antiproton flux from DM annihilation  
is dominated by local contributions.

\item The impact of different DM density distribution is relevant considering 
the antiproton bound obtained using the KOL, THK and KRA
propagation models. In these models, therefore, a large contribution on the antiproton flux from DM annihilation comes from non-local regions pointing towards the Galactic center where the three profiles present sizable differences, being  more or less peaked. In greater detail, we find 
that the Isothermal (Einasto) profile gives the weaker (stronger) constraint;
moreover, comparing with the NFW case, 
 the bound obtained 
assuming the Isothermal density distribution has the largest deviation, while the Einasto density distribution gives a similar result.
Comparing the three profiles, as done in fig.~\ref{fig:profiles},
we notice that in the region  $r\gtrsim 0.5$ kpc -- being r the radial distance from the Galactic center --
the density distribution in both the Einasto and the NFW profiles are significantly larger than the Isothermal case;
moreover, in this region the Einasto density distribution is larger if compared with the NFW one, thus reflecting the
hierarchy observed in the exclusion curves.
 On the contrary, for $r\lesssim 0.5$ kpc, the density distribution in the NFW case is larger w.r.t. the
  Einasto profile. All in all, we argue that in the KOL, THK and KRA
propagation model the antiproton flux from DM annihilation is dominated by regions close to the Galactic center,  
with $0.5 \lesssim r \lesssim r_{\odot}$ kpc. 
In order to strengthen this argument, we show in fig.~\ref{fig:RadialDependence} the local antiproton flux coming from DM annihilation and, in the inset plot,
 the relative contribution from a region enclosed within $1$ kpc from the Galactic center. For definiteness, we consider  $\langle \sigma v_{\rm rel}\rangle_{0} = 3\times 10^{-26}$ cm$^3$s$^{-1}$ and DM annihilation into $\bar{b}b$ ($W^+W^-$)
for $m_{\rm S} = 70$ GeV ($m_{\rm S} = 700$ GeV).\footnote{Without solar modulation, as plotted in the left panel of fig.~\ref{fig:RadialDependence}, the CON and THN models have similar flux at given DM mass ($m_S = 70$ GeV) and cross section. 
However, their antiproton bounds are different as shown in fig.~\ref{fig:ScalarPortalLowMass}. The reason is that
the CON model gives harder cosmic-ray spectrum, which asks for smaller value of solar modulation. 
Taking properly into account the different values of solar modulation, the THN model has looser bound than the CON one.}   
From this plot it is clear that for all the propagation models the contribution to the total antiproton flux from the inner Galactic region is at most $20$\%. Moreover, as expected, the THN and CON propagation models receive a negligible contribution from the region with $r < 1$ kpc. The thin height of the diffusion zone and the presence of strong convective wind efficiently remove 
a large fraction of the antiprotons originated towards the center of the Galaxy increasing their escape probability.  
 
\end{itemize}

In conclusion, we have shown that the role of the DM density distribution
in the Galaxy  plays only a relatively marginal role in our analysis, 
and  the astrophysical uncertainties affecting the limits on the scalar Higgs portal model that 
one can obtain using the antiproton data are mostly dominated by 
the details of the propagation model.\footnote{We remind that in 
this paper we do not consider additional uncertainties that could affect the computation of the antiproton flux both from DM annihilation and cosmic rays. 
Strong outflows from the Galactic center, for instance, can carry away a lot of annihilation products thus weakening the correlation between the $\bar{p}$ locally observed at Earth and the ones produced by DM annihilation; a more detailed analysis of astrophysical uncertainties can be found in ref.~\cite{Evoli:2011id}. Notice, moreover, that in our analysis we extract the bound on DM using a fixed value for the solar modulation potential (the one obtained from the best-fit of the background contribution, see table~\ref{tab:models}). A different approach is based on the possibility to treat this variable as a nuisance parameter in the fit, thus leading to slightly weaker constraints; a more detailed analysis along this line can be found in refs.~\cite{Fornengo:2013xda,Cirelli:2014lwa}. 
Finally, there is an additional source of uncertainty related to the nuclear cross-sections describing the production of secondary $\bar{p}$. If introduced, this additional uncertainty will weaken the bound \cite{Fornengo:2013xda}.} In this regards, 
the antiproton data that will be released by the AMS-02 experiment will play a 
crucial role in order to improve the current sensitivity.
In the next section, therefore, we will briefly discuss future perspectives.

\section{Future perspectives}\label{sec:future}

In this section we discuss some future perspectives related to our analysis.
In section~\ref{sec:AMS02}, we analyze the constraining power of the antiproton data
that will be released by the AMS-02
experiment. In section~\ref{sec:Antideuteron}, we analyze the antideuteron flux from DM annihilation in the scalar Higgs portal model.

\subsection{AMS-02}\label{sec:AMS02} 

The Alpha Magnetic Spectrometer (AMS-02) is a particle physics detector 
hosted on board of the International Space Shuttle, and designed to measure various cosmic-ray fluxes;
thanks to this instrument, in the next future a more precise determination of the antiproton and antiproton-to-proton ratio will improve the constraints  derived in this paper. 
To get a more concrete idea, we can simulate the prospects of this experiment by means of a set of mock data,
in the energy range of (1 GeV, 450 GeV).
Following ref.~\cite{AMS02slides}, we use a linear approximation of the AMS-02 detector energy 
resolution 
\begin{equation}
  \Delta E / E = \left( 0.042~\frac{E}{{\rm GeV}}    + 10 \right) \%  \ ,
\end{equation}
 \cite{DeSimone:2013fia,Cirelli:2013hv}. 
which determines the energy bin-size of the data. 
Having the bin size $\Delta E_i$ and the flux $\Phi_i$ for each bin, the observed 
antiproton number can be derived as $N_{\bar{p}}=\epsilon~a_{\bar{p}}~\Phi_i~\Delta E_i~\Delta t$, where
we take $\epsilon \simeq 1$ for the  efficiency, 
$a_{\bar{p}} = 0.2$ m$^2$sr for the geometrical acceptance of the instrument, and  $\Delta t = 1$ year for the reference 
data taking time.
Since the dominant statistical uncertainty comes from the antiproton rather than the proton flux, the statistical error is approximately $1 /\sqrt{N_{\bar{p}}}$; 
for definiteness, we fixed the systematic uncertainties to be $5\%$ for one-year data taking. The uncertainty at each data point is the sum in quadrature of 
the systematic and statistical errors. The central value of the data for each bin, $\Phi_i$,  follows the predicted 
flux from our five benchmark propagation models, which does not contain any DM 
contribution. With these mock data in hand, we can study the future sensitivity of antiproton-to-proton ratio data 
on DM models.

\begin{figure*}[!htb]
\minipage{0.5\textwidth}
  \includegraphics[width=\linewidth]{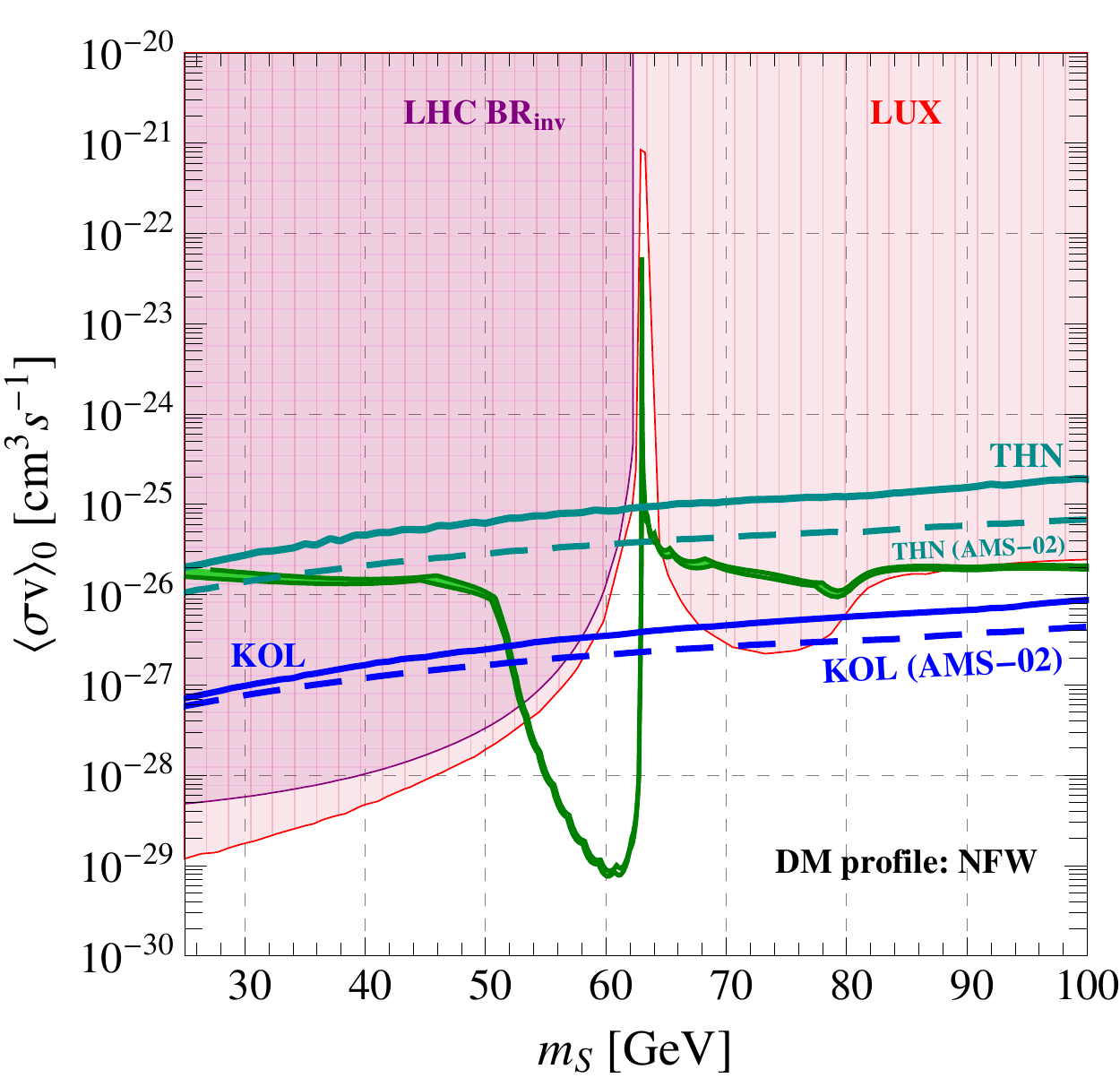}
\endminipage
\minipage{0.5\textwidth}
  \includegraphics[width=\linewidth]{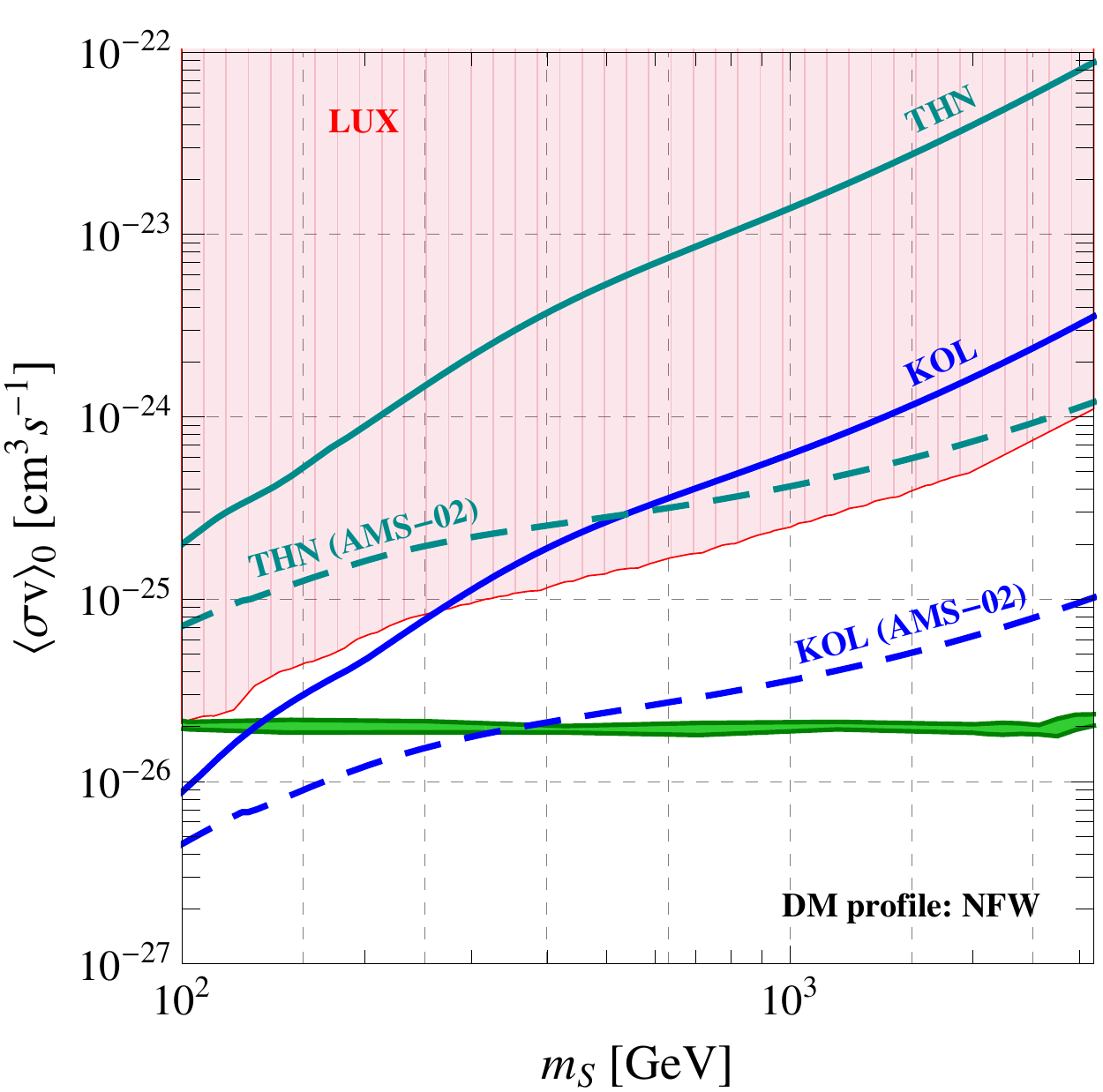}
\endminipage
 \caption{\textit{
Projected bounds on the annihilation cross-section at zero temperature in the low- (left panel) and high-mass region (right panel) considering the THN (dark cyan) and KOL (blue) propagation models. Solid lines represent the current bounds obtained in 
section~\ref{sec:Results} while dashed lines are the projected bounds obtained considering mock data for the AMS-02 experiment. The green line matches the observed value of DM relic abundance.
 }}\label{fig:ScalarPortalMockAMS}
\end{figure*}

In fig.~\ref{fig:ScalarPortalMockAMS} we show 
the projected bounds on the DM annihilation cross-section in the low- (left panel) and high-mass region (right panel) considering, as two extreme cases, the THN and KOL propagation models. We find that the future AMS-02 antiproton data will improve the bound for more than an order of magnitude in the high-mass region.
On the contrary, in the low-mass region, our analysis show a little improvement if compared with the existing data. 
This is because we already exploited the full available data set, which is of reasonably good quality in the low-energy region, and goes until the lowest value of $0.1$ GeV, while the mock AMS-02 data starts from $1$ GeV.
In the high-energy 
region, on the contrary, AMS-02's resolution and luminosity are much better than the current status.

 \subsection{Antideuteron}\label{sec:Antideuteron}
 
Antideuteron  has been proposed in ref.~\cite{Donato:1999gy}
as a promising indirect signal of DM annihilation in the Galactic halo.
On a general ground, the annihilation of DM particles into SM hadronic channels -- i.e. $q\overline{q}$, $W^+W^-$, and
 $ZZ$ -- may produce antideuterons in the final state as a consequence of a two-step 
process. First -- after showering, hadronization, and decay of unstable particles -- a large number of antiprotons 
and antineutrons are produced. Second, an antiproton-antineutron pair may coalesce
 to form an antideuteron nucleus. The description of this process is usually addressed in the context of 
 the so-called coalescence model. Given an antiproton and an antineutron with 
 four-momenta $k_{\bar{p}}^{\mu}$ and $k_{\bar{n}}^{\mu}$, the coalesce model 
 approximates the probability for the formation of an antideuteron 
 with the step function $\Theta(\Delta^2 + p_0^2)$, where $\Delta^{\mu} \equiv 
 k_{\bar{p}}^{\mu} - k_{\bar{n}}^{\mu}$, and $p_0$ is the maximum value of relative momentum that 
 allows to form an antideuteron bound state. 
 In this picture, $p_0$ is a free parameter and 
its numerical value has to be extracted from experimental data (see refs.~\cite{Ibarra:2012cc,Fornengo:2013osa} for a detailed discussion). 
In our analysis we use the results of ref.~\cite{Cirelli:2010xx} in order to reconstruct the antideuteron 
energy spectra
produced by DM annihilation. These energy spectra have been obtained using $p_0 = 160$ MeV, and
the coalescence model previously discussed has been applied studying DM annihilations event-by-event \cite{Kadastik:2009ts}. 
Let us now move to discuss the antideuteron produced by high-energy astrophysical phenomena.
The most relevant argument supporting the claim in ref.~\cite{Donato:1999gy}, in fact,
 emerges from the comparison between
 the antideuteron signal produced by DM annihilation and the corresponding astrophysical background.
 To be more precise, there are two key points to keep in mind.
\textit{i})~Antideuterons are produced by high-energy collisions between extragalactic 
 cosmic rays (mostly $p$, $\bar{p}$ and He) and the interstellar gas (mostly H and He) in the Galactic disk; the corresponding cross-section is small and -- most importantly -- it is characterized by a relatively high kinematical threshold; for instance the energy threshold for the creation of an antideuteron from a collision of a cosmic ray proton (antiproton)
 with the interstellar gas is $E_{\rm th} = 17$ m$_p$ ($E_{\rm th} = 7$ m$_p$).
Let us give a closer look to this numbers considering for definiteness the scattering between a cosmic ray proton and a proton at rest in the interstellar gas.
  Because of conservation of baryon number, the production of an antideuteron 
 from a proton-proton collision requires a six-body final state, with a total energy square
  $\hat{s}>(6m_p)^2$. On the other hand, in the rest frame of the gas, $\hat{s} = (m_p + E_{p})^2 - k_{p}^2$,
  where $E_p$ and $k_p$ are the energy and momentum of the cosmic ray proton. 
  From these considerations, it follows that the impinging 
  cosmic ray proton needs to have an energy $E_p > 17$ m$_p$ in order to create an antideuteron. \textit{ii}) The binding energy for an antideuteron
  is extremely low, namely $B_{\bar{d}} \approx 2.2$ MeV. This implies that antideuterons 
  are easily destroyed, and -- as a consequence -- they do not have to possibility to 
  propagate long enough in order to loose most of their energy. 
  The astrophysical background of antideuterons with kinetic energy $E_{\bar{d}} \lesssim 1-3$ GeV, therefore, is expected to be extremely small. Below these kinetic energies, an antideuteron flux originated from
  DM annihilation may easily stand out from the astrophysical background for more than one order of magnitude.
\begin{figure*}[!htb!]
\minipage{0.5\textwidth}
  \includegraphics[width=\linewidth]{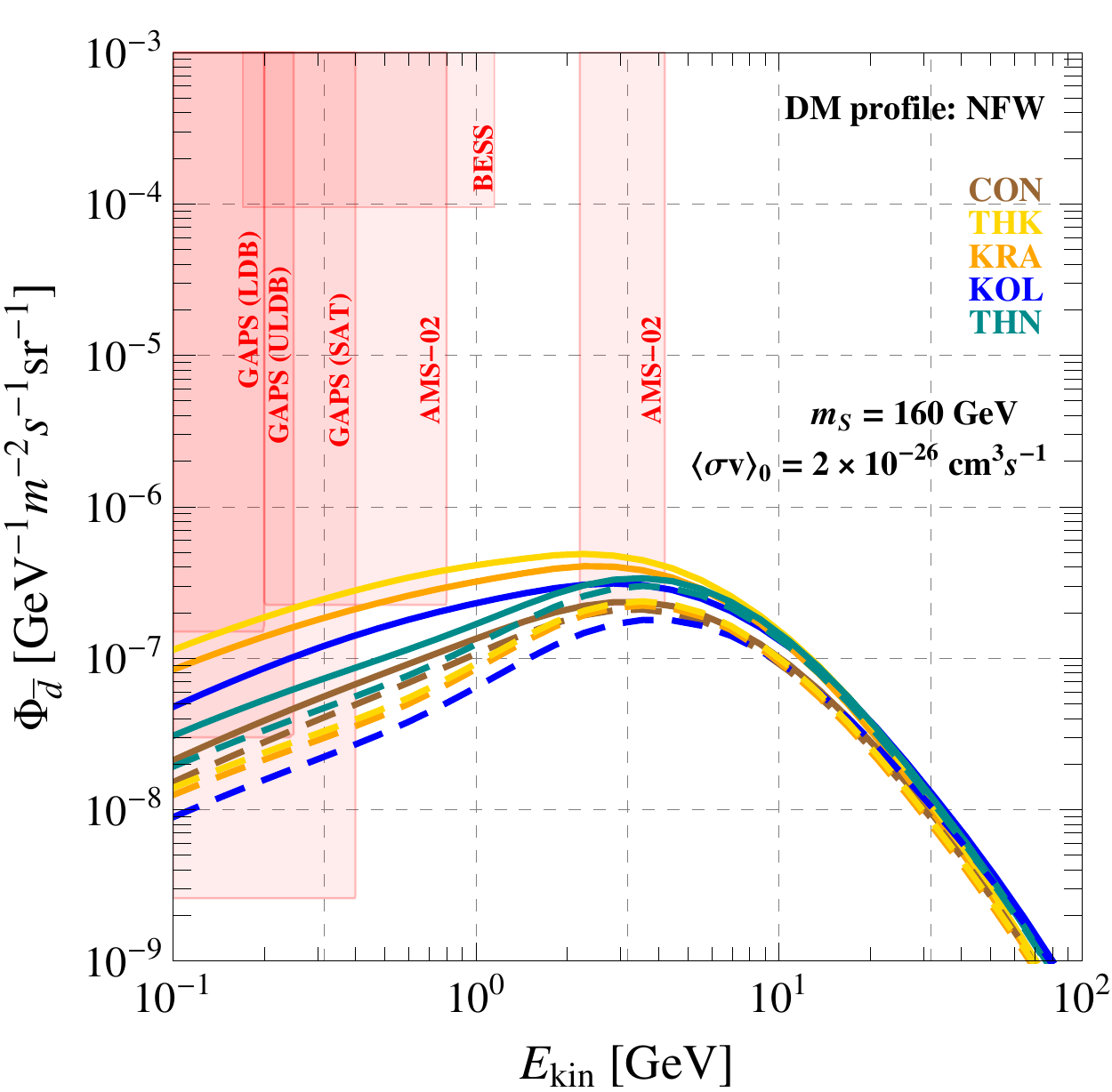}
\endminipage
\minipage{0.5\textwidth}
  \includegraphics[width=\linewidth]{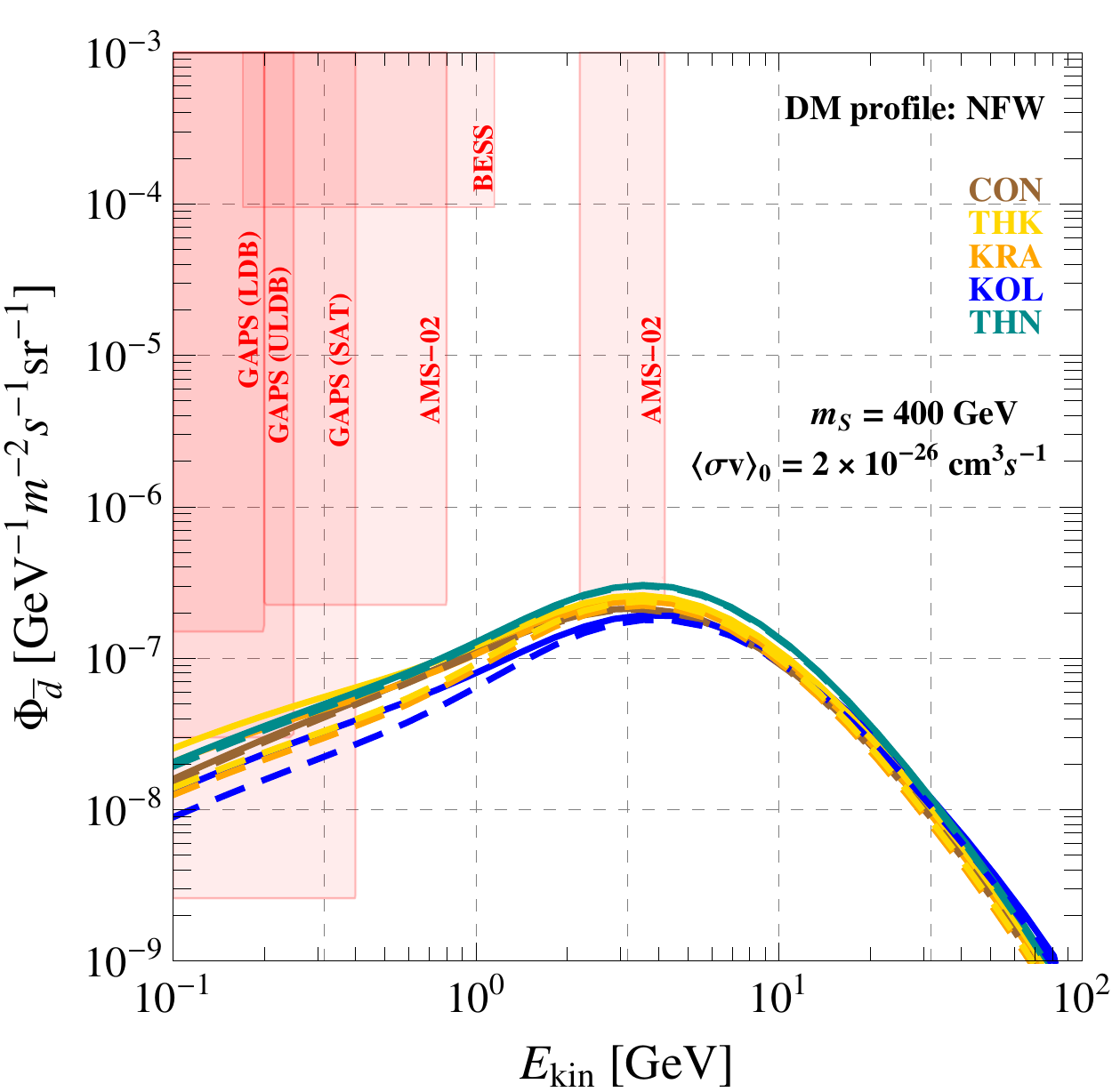}
\endminipage
 \caption{\textit{
Antideuteron  fluxes for $m_{\rm S}=160$ GeV (left panel) and  $m_{\rm S}=400$ GeV  (right panel), with $\langle \sigma v_{\rm rel}\rangle = 2\times 10^{-26}$ cm$^3$s$^{-1}$.
Dashed lines correspond to the background while 
solid lines to the background plus DM signal. We show the results obtained using the five propagation models in table~\ref{tab:models}. The shaded regions give the exclusion by BESS experiment and projected sensitivity
of GAPS and AMS-02 experiments (see, e.g., ref.~\cite{Hryczuk:2014hpa}). 
As a reference value, we fixed the solar modulation potential to be $\Phi = 0.25$ GV.
 }}\label{fig:Antideuteron}
\end{figure*}
In order to translate these qualitative statements into more
quantitative results,  
we need to compare antideuteron background and DM signal after propagation in the Galaxy.
 The cross-sections describing production, elastic and inelastic scattering of antideuteron -- key ingredients in order to solve the corresponding propagation equation, see eq.~(\ref{eqn::prop}) -- are not well known.
 Following ref.~\cite{Hryczuk:2014hpa}, we implemented in the DRAGON code
 all these cross-sections
 using the results of ref.~\cite{Duperray:2005si}, where they were extrapolated from experimental data under reasonable assumptions. In this way, we are in the position to solve numerically the propagation equation for antideuterons 
 considering both the astrophysical background and the DM signal.
 
 Present and future experiments -- two energy bands of the AMS-02 experiment and the three phases of the General AntiParticle Spectrometer (GAPS) \cite{Mori:2001dv,Fuke:2008zz} -- will increase the sensitivity of searches 
 for cosmic-ray antideuteron over the current bound set by the BESS experiment. For the 
 proposed sensitivities of AMS-02 and GAPS experiments we use the values from ref.~\cite{Hryczuk:2014hpa}.
 In fig.~\ref{fig:Antideuteron}, we present the predictions of the scalar Higgs portal model for the antideuteron flux, 
 comparing background and background plus DM signal hypothesis. 
 We analyze two benchmark values for the DM mass, and we use for definiteness the NFW density profile. In the left panel of fig.~\ref{fig:HiggsGamma} we take 
 $m_{\rm S} = 160$ GeV, with $\langle \sigma v_{\rm rel}\rangle_0 = 2\times 10^{-26}$ cm$^3$s$^{-1}$.
 These values correspond to a DM candidate that, according to figs.~\ref{fig:ScalarPortalHighMass},~\ref{fig:CombinedPlotHighMass}, lies close to the present bound 
 placed by the analysis of the PAMELA antiproton data. 
 We find that the corresponding total antideuteron flux  
is higher than all the experimental sensitivities  of the  GAPS experiment assuming the KOL, THK and KRA propagation models.  Therefore in these cases, if such DM candidate is realized in Nature, we expect -- in principle -- a combined detection in both antideuteron and antiproton  channels. For illustrative purposes, we show in the right panel of 
fig.~\ref{fig:Antideuteron} a different situation with $m_{\rm S} = 400$ GeV, $\langle \sigma v_{\rm rel}\rangle_0 = 2\times 10^{-26}$ cm$^3$s$^{-1}$. 
These values correspond to a DM candidate that, according to fig.~\ref{fig:ScalarPortalMockAMS}, lies close to the future bound 
 that will be placed by the analysis of the AMS-02 antiproton data. 
 In this case, however,
 the total antideuteron flux will be hardly distinguishable from the astrophysical background.

\section{Conclusions and outlook}\label{sec:Conclusions}

In this work, we extracted a new bound on the scalar Higgs portal DM model 
using high-energy cosmic-ray astrophysics. In summary, the main points of our analysis are the following.
First, we studied the propagation equation that governs the motion of charged particles 
in the Milky Way galaxy using the numerical code DRAGON. This equation depends on several free parameters that encode the astrophysical uncertainties describing the propagation of cosmic rays in the Galaxy. We fixed some of these parameters -- i.e. the halo thickness, the source spectral index,  the rigidity slope and the gradient of the convection velocity in the vertical direction -- so to define five different propagation setups.
Using the measurement of the boron-to-carbon ratio performed by the HEAO-3, ACE, CREAM, ATIC and CRN experiments, we fixed, via a minimization procedure, the remaining ones -- i.e. the normalization of the diffusion coefficient, the Alfv\'en velocity and the low-energy diffusion index. 
Second, using the propagation setups fixed by the B/C data, we predicted the background contribution to the antiproton-to-proton ratio. Finally, we computed the antiproton flux from DM annihilation in the Higgs portal model including three-body final states and QCD radiative corrections; using the antiproton-to-proton ratio previously discussed, and combining 
background and DM signal, we extracted 3-$\sigma$ bound on the parameter space of the model. 
The use of the antiproton-to-proton ratio allowed us to combine different experiments, namely PAMELA, BESS and CAPRICE data. In the antiproton-to-proton ratio, in fact, several systematic effects that plague the comparison between different experiments -- as for instance different energy calibration -- are integrated out. 
We compared our antiproton bound with the constraints coming from the LUX and LHC experiments considering -- respectively -- direct detection of DM particles and the invisible decay width of the Higgs boson. At the same time, we required to reproduce the observed amount of relic density. 

We found that the antiproton bound is competitive; in particular, it provides the most stringent constraint on the model in the mass range $m_{\rm S} \approx 80-300$ GeV  for most of the analyzed propagation setups. Most importantly, the antiproton bound 
is the only one able to put in significant tension the resonant region $m_{\rm S}\approx m_h/2$, otherwise of difficult access to direct detection and collider searches.

 In our analysis, we investigated the impact of astrophysical uncertainties related to different propagation setups and different models for the DM density distribution in the Galaxy. Moreover, we discussed future perspectives using a set of mock data in order to simulate those that will be released  in the near future by the AMS-02 experiment. Finally, we highlighted in the context of the scalar Higgs portal model the role of the antideuteron channel
as an important indirect detection observable able to provide a signature of annihilating DM.

\acknowledgments{

We thank Daniele Gaggero, Andrzej Hryczuk and in particular Piero Ullio for useful discussions and advices. The work of A.U. is supported by the ERC Advanced Grant n$^{\circ}$ $267985$, ``Electroweak Symmetry Breaking, Flavour and Dark Matter: One Solution for Three Mysteries" (DaMeSyFla).}

\begin{appendix}

\section{Spin-1/2 Higgs portal}\label{sec:FermionHiggsPortal}
In this Appendix we focus on the following fermionic Higgs portal Lagrangian \cite{LopezHonorez:2012kv}
\begin{equation}
\mathcal{L}_{\rm fHP} = \mathcal{L}_{\rm SM}
+\bar{\chi}(i\slashed{\partial}-m_{\chi})\chi
+\frac{d_{\chi}}{\Lambda}|H|^2\bar{\chi}\chi
+\frac{ic_{\chi}}{\Lambda}|H|^2\bar{\chi}\gamma^5\chi~,
\end{equation}
where $\chi$ is Dirac field playing the role of DM. The parity-conserving interaction $d_{\chi}$ is severely constrained by direct detection experiments \cite{Kanemura:2010sh}; moreover the annihilation cross-section suffers from a p-wave suppression (see, e.g., Ref.~\cite{Huang:2013apa})
that makes it undetectable for any indirect detection experiment. The bound discussed in this paper, therefore, does not apply on this interaction. The parity-violating interaction $c_{\chi}$, on the contrary, induces a velocity-suppressed spin-independent elastic cross-section on nuclei but an unsuppressed annihilation cross-section. 
As a consequence we expect that the bound from antiprotons, 
in absence of a significant direct detection signal, is the strongest constraint that can be placed on the parameter space of this model. 

Compared with eq.~(\ref{eq:Xsection}), the annihilation cross-section times relative velocity is
 \begin{equation}\label{eq:FermionXsection}
\sigma v_{\rm rel}=\frac{1}{\sqrt{s}}\left[\frac{s\lambda_{\chi}^2 v^2}{(s-m_h^2)^2+\Gamma_{\rm h,\chi}^2 m_h^2}\right]\Gamma_{\rm h}(\sqrt{s})~,
\end{equation}
with $\lambda_{\chi}\equiv c_{\chi}/\Lambda$, $\Gamma_{\rm h,\chi}\equiv \Gamma_{\rm h}^{\rm SM}+\Gamma_{\rm h\to \bar{\chi}\chi}$ and 
\begin{equation}
\Gamma_{\rm h\to \bar{\chi}\chi}(m_{\chi}, \lambda_{\chi}) 
= \frac{m_{h} v^2\lambda_{\chi}^2}{8\pi}
\sqrt{1-\frac{4m_{\rm S}^2}{m_h^2}}~.
\end{equation}
The phenomenological analysis proceeds parallel to the scalar case.

\begin{figure*}[!htb]
\minipage{0.5\textwidth}
  \includegraphics[width=\linewidth]{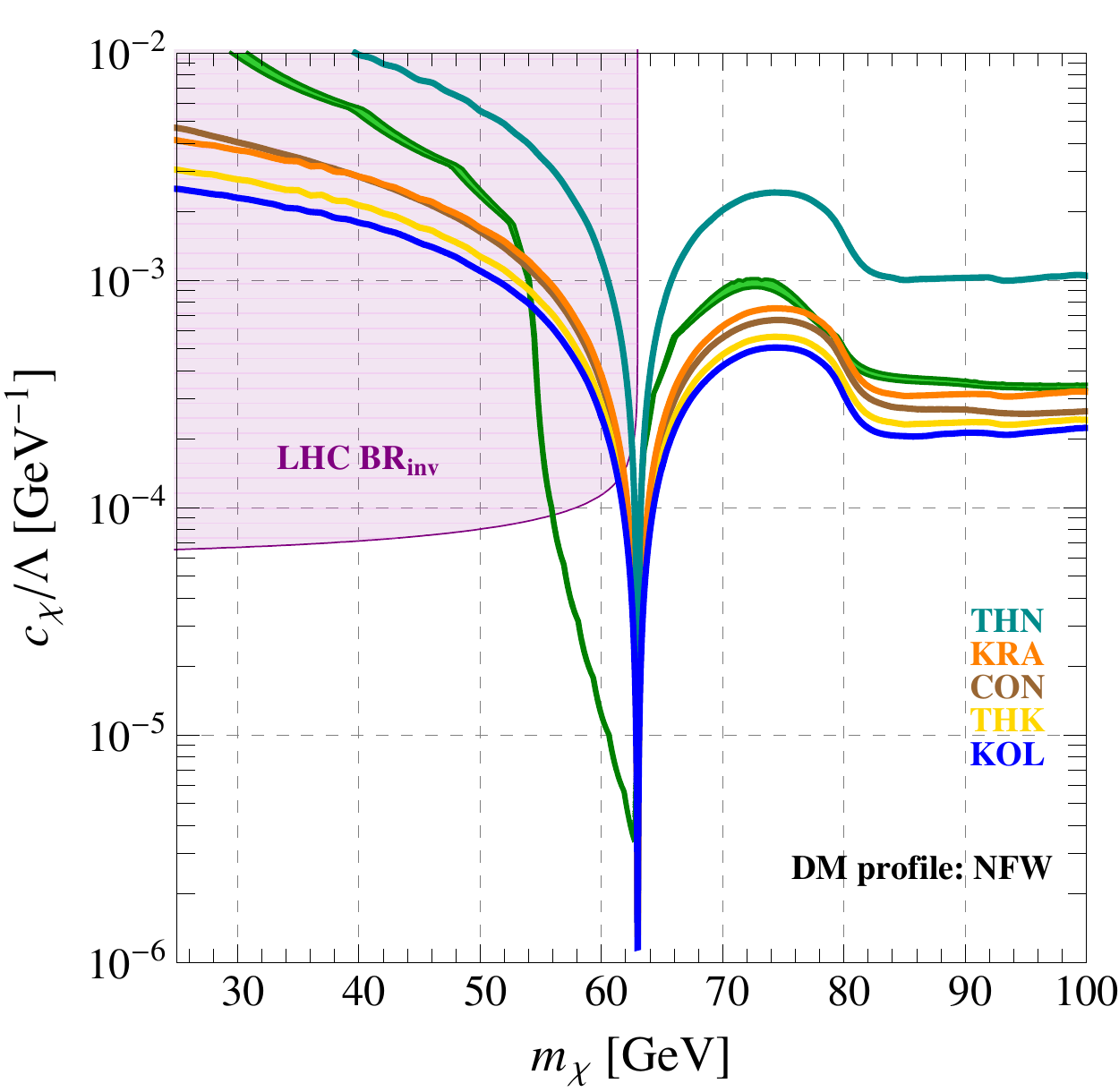}
\endminipage
\minipage{0.5\textwidth}
  \includegraphics[width=\linewidth]{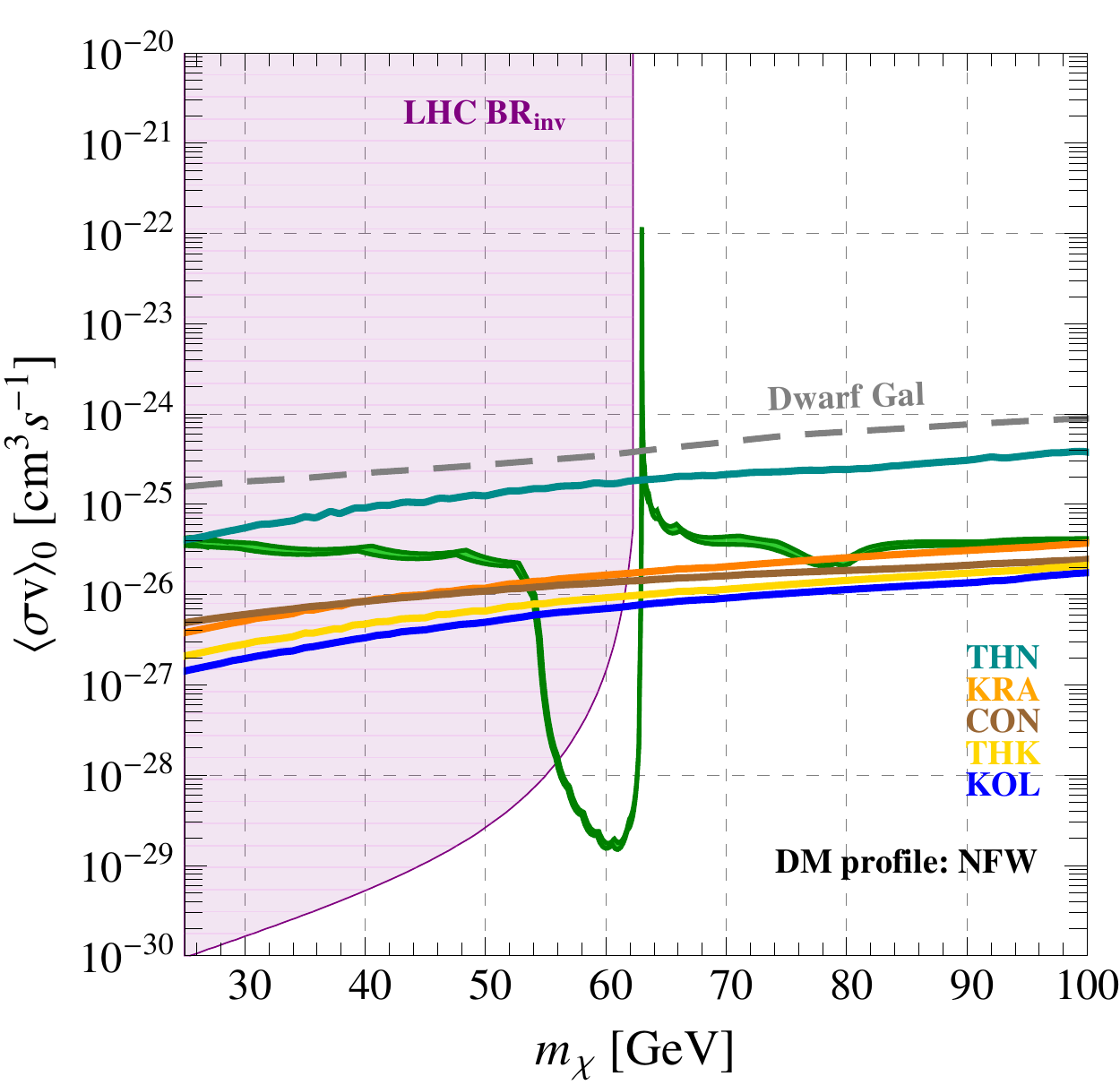}
\endminipage
 \caption{\textit{
 Same as in Fig.~\ref{fig:ScalarPortalLowMass} but for the fermionic Higgs portal described in Section~\ref{sec:FermionHiggsPortal}.
 }}\label{fig:FermionPortalCombo}
\end{figure*}

\end{appendix}


\bibliography{ref}
\bibliographystyle{jhep}

\end{document}